\documentclass{ws-ijmpd}

\begin{document}

\markboth{R. Foot}{Mirror matter-type dark
matter}

\catchline{}{}{}{}{}

\title{Mirror matter-type dark matter }

\author{R. Foot}
\address{
rfoot@unimelb.edu.au \\
School of Physics, University of Melbourne, 3010 Australia. }

%%%%%%%%%%%%%%%%%%%%%%%%%%%%%%%%%%%%%%%%%%%%%%%%%%%%%%%%%%%%
% You may repeat \author \address as often as necessary    %
%%%%%%%%%%%%%%%%%%%%%%%%%%%%%%%%%%%%%%%%%%%%%%%%%%%%%%%%%%%%

\maketitle

%\begin{history}
%\received{(DAY MONTH YEAR)}
%\revised{(DAY MONTH YEAR)}
%\end{history}

\begin{abstract}
There are six main things which any non-baryonic dark matter
theory should endeavour to explain: 
(1) The basic dark matter particle properties [mass, stability, darkness];
(2) The similarity in cosmic abundance between ordinary
and non-baryonic dark matter, $\Omega_B \sim \Omega_{dark}$;
(3) Large scale structure formation; 
(4) Microlensing (MACHO) events;
(5) Asymptotically flat rotation curves in spiral galaxies; 
(6) The impressive DAMA/NaI annual modulation signal.
Only mirror matter-type dark matter is capable of 
explaining all six of these desirable features. The purpose
of this article is to provide an up-to-date 
and pedagogical review of this dark matter candidate.

\end{abstract}


%%%%%%%%%%%%%%%%%%%%%%%%%%%%%%%%%%%%%%%%%%%%%%%%%%%%%%%%%%%%
% The main text of your paper	begins here			     %
%%%%%%%%%%%%%%%%%%%%%%%%%%%%%%%%%%%%%%%%%%%%%%%%%%%%%%%%%%%%

\section{Introduction}

There is a very strong scientific case that most
of the matter in the Universe consists of non-baryonic
stable particles. Since the standard model of particle
physics does not contain any heavy stable non-baryonic
particles new particle physics is required. But 
what is this new physics?

It is widely assumed that the
particles comprising non-baryonic dark matter
are weakly interacting in the sense that they interact
with ordinary matter via exchange of
W, Z gauge bosons, Higgs bosons or more exotic heavy
particles. From collider bounds (e.g.
lack of new particles in decays of the W and Z gauge bosons),
the masses of any new weakly interacting particles should
be (typically) greater than about 30-45 GeV\footnote{
Throughout this article we use units where $\hbar = c = k = 1$, unless
indicated otherwise.} -- depending on the model. 
However, such heavy
weakly interacting particles should decay with a lifetime
of order $\sim 1/(g^2 M_{wimp}) \sim 10^{-24}$ seconds (for $M_{wimp}
\sim
M_Z$) -- about 41 orders of magnitude
too short-lived to be suitable as dark matter candidates\footnote{
E.g. the Z-boson is a weakly interacting massive particle
--  a WIMP -- and has lifetime of about $3\times 10^{-25}$ seconds.}.
Thus, one must make additional ad hoc assumptions in order
to stabalize such hypothetical particles.
In the end, theories with such particles require multiple
unrelated assumptions and become very ugly from a particle
physics point of view.
A well known example is the popular neutralino model
which requires at least three independent hypothesis:
a) broken low energy supersymmetry exists which provides
WIMP candidates, b) an exact unbroken r-parity
symmetry is proposed to prevent the lightest
superpartner from decaying\footnote{Note that r-parity has, 
of course, nothing to do with
space-time parity but is an ad hoc discrete symmetry.} 
and c) the lightest
superpartner is hypothesised to be neutral to make it
suitable for dark matter.

It seems to me that a more plausible candidate for
this new physics arises from the hypothesis of
exact unbroken mirror symmetry [$x \to -x, \ t \to t$].
It is more plausible because it involves
only a single well motivated hypothesis. 
Improper space-time symmetries, such as
parity and time reversal symmetries, stand out as the 
only obvious symmetries which are not respected by
the interactions of the known elementary particles. It is an interesting
and non-trivial fact that these symmetries can
be exact, unbroken symmetries of nature if a 
set of mirror particles exist. 
Even more interesting is that the mirror particles
have the right broad properties to be identified with the non-baryonic
dark matter in the Universe.

The ordinary and mirror particles form parallel sectors each
with gauge symmetry $G$ (where $G=G_{SM} \equiv SU(3)_c \otimes SU(2)_L 
\otimes U(1)_Y$
in the simplest case) so that the full gauge group is $G \otimes G$.
Mathematically, mirror symmetry has the form:\cite{flv}
\begin{eqnarray}
& x \to -x, \ t \to t, \nonumber \\
& 
G^{\mu} \leftrightarrow G'_{\mu},\ 
W^{\mu} \leftrightarrow W'_{\mu}, \ B^{\mu} \leftrightarrow B'_{\mu},
\nonumber \\
& \ell_{iL} \leftrightarrow \gamma_0 \ell'_{iR}, \
e_{iR} \leftrightarrow \gamma_0 e'_{iL}, \
q_{iL} \leftrightarrow \gamma_0 q'_{iR}, \
u_{iR} \leftrightarrow \gamma_0 u'_{iL}, \
d_{iR} \leftrightarrow \gamma_0 d'_{iL}, 
\end{eqnarray}
where $G^{\mu}, W^{\mu}, B^{\mu}$ are the standard
$G_{SM} \equiv SU(3)_c \otimes SU(2)_L \otimes U(1)_Y$ gauge particles,
$\ell_{iL}, e_{iR}, q_{iL}, u_{iR}, d_{iR}$ are the
standard leptons and quarks ($i=1,2,3$ is the generation index)
and the primes denote the mirror particles. There is also a 
standard Higgs doublet $\phi$ with a mirror Higgs doublet partner,
$\phi'$. Importantly, there is a large range of
parameters of the Higgs potential for which
mirror symmetry is {\it not} spontaneously
broken by the vacuum (i.e. $\langle \phi \rangle = \langle \phi'
\rangle$) so that it is an exact, unbroken
symmetry of the theory\cite{flv}.
\footnote{
It is theoretically possible to have mirror 
symmetry spontaneously broken by the vacuum ($\langle \phi \rangle
\neq \langle \phi' \rangle$), but the simplest models
of this type are disfavoured for a variety of reasons\cite{flv5}.
[More complicated models with broken
mirror symmetry are still possible and have been studied
in the literature, see e.g. ref.\cite{complicated}].}
Interestingly, despite doubling the
number of particle types the number of free parameters have
not (yet!) been increased; mirror symmetry implies that the masses and 
couplings of the 
particles in the mirror sector are exactly the same as the 
corresponding ones in the ordinary sector.

Ordinary and mirror particles couple with each other via gravity
and possibly by new interactions connecting ordinary and mirror
particles together. 
Constraints from gauge invariance, mirror symmetry and
renormalizability, suggest only two
types of new interactions\cite{flv}:
a) Higgs-mirror Higgs quartic coupling
(${\cal L} = \lambda' \phi'^{\dagger}\phi' \phi^{\dagger} \phi$),
and b) via photon-mirror photon kinetic mixing:
\footnote{
Allowing the ordinary and mirror sectors to interact with each other leads to 
two new free parameters ($\lambda', \epsilon$).  However, compared to
other ideas beyond the standard model, many of which have
literally hundreds of new parameters, mirror symmetry {\it is}
a fairly minimal extension of the standard model.
Also note, if the neutrinos have mass, mass mixing between ordinary and
mirror neutrinos is also possible\cite{flv2,f94} and might be
implicated by the observed atmospheric, solar and 
LSND neutrino anomalies. 
However, the experimental situation is still not clear\cite{foot}.
}
\begin{eqnarray}
{\cal L}_{int} = {\epsilon \over 2}F^{\mu \nu}F_{\mu \nu}' ,
\label{km}
\end{eqnarray}
where $F^{\mu \nu}$ ($F'_{\mu \nu}$)
is the field strength tensor for electromagnetism (mirror
electromagnetism)
\footnote{
Technically, the photon-mirror photon kinetic mixing arises from
kinetic mixing of $U(1)_Y, U(1)'_Y$ gauge 
bosons, since only for abelian $U(1)$ symmetry is the mixing
term, $FF'$, gauge invariant\cite{fh}.
Therefore there is both $\gamma-\gamma'$ and  $Z-Z'$
kinetic mixing. [However, experiments are much more sensitive to
$\gamma - \gamma'$ kinetic mixing which is why it is more
important].
In the case of theories without $U(1)$ gauge symmetries, such
as GUTs, the $\gamma-\gamma'$ mixing can arise radiatively
provided that there exists a mixed form of matter carrying both ordinary and mirror electric
charges\cite{hol}.
Interestingly, there is a class of such models\cite{cf} where $\epsilon$
vanishes at one and two loop level, and therefore
naturally of the order of $\epsilon \sim 10^{-8}$.}. 
The effect of the Higgs-mirror Higgs quartic coupling is to
modify the properties of the standard Higgs boson\cite{flv,flv2,sasha}. This
interaction will be tested if/when scalar particles are
discovered. One effect of photon-mirror photon kinetic mixing is to
cause mirror charged particles
(such as the mirror proton and mirror electron)
to couple to ordinary photons with effective electric
charge $\epsilon e$\cite{flv,hol,s}.
As we will see, this 
non-gravitational interaction between ordinary
and mirror particles provides a key way to experimentally test
the theory.

To summarize:  the only obvious space-time symmetries that are
not respected by the interactions of the known elementary
particles are the improper Lorentz symmetries (such
as parity and time reversal). 
These symmetries can be unbroken 
symmetries of nature provided that the
Universe contains
both ordinary and mirror particles.
The mirror particles have identical masses to the corresponding
ordinary particles
and have identical, but separate gauge interactions (the gauge
group is $G_{SM} \otimes G_{SM}$).
The mirror particles couple to the ordinary ones via gravity and
possibly via Higgs-mirror Higgs interactions and photon-mirror
photon kinetic mixing.
It turns out that the mirror
particles lead to an elegant explanation for
the inferred non-baryonic dark matter component of the Universe,
as we will explain in more detail in the following sections.

\vskip 0.3cm
\section{Identifying mirror matter with the inferred non-baryonic dark matter
in the Universe}
\vskip 0.3cm

\noindent
There is a substantial range of evidence for non-baryonic
dark matter in the Universe. We have already emphasised in
the introduction that a basic requirement is that the 
massive particles comprising dark matter need to be stable and
have no or small coupling to ordinary photons. 
There are other desirable features that are also
required, including (in random order):
\begin{itemize}
\item 
An explanation for $\Omega_{dark} \sim \Omega_B$.

\item
It should be capable of explaining the large scale
structure of the Universe. 

\item
Asymptotically flat rotation curves in spiral galaxies suggest 
that dark matter
is (roughly) spherically distributed in a `halo' in spiral galaxies.
This is in contrast to ordinary matter which is distributed in
the disk and bulge. 

\item
A substantial fraction ($\sim 20\%$) of the mass of the halo appears to be
in the form of massive ($\sim 0.5 M_{\odot}$) 
compact invisible objects (MACHOs).
What are the MACHOs and why are they invisible?

\item
The direct experimental detection of halo dark matter particles has
been achieved by the DAMA/NaI collaboration. Other experiments
report only negative results. Why?
\end{itemize}

\noindent
We now examine each of these items from a mirror matter
perspective.

\subsection{$\Omega_{dark} \sim \Omega_B$}

Precision cosmic microwave background measurements (culminating
with the recent WMAP results\cite{cmb}) have established that the
Universe is spatially flat, i.e. $\Omega_{tot} \simeq 1.0$.
Furthermore, the WMAP results, together with observations of
high redshift Type 1a supernovae\cite{hr}, 
and other measurements, suggest that
the Universe consists predominately of three components:
ordinary matter ($\Omega_B \approx 0.05$),
non-baryonic dark matter ($\Omega_{dark} \approx 0.22$) and 
dark energy ($\Omega_{\Lambda} \approx 0.7$).
It is striking that each of these three different components
should have energy densities
of the same order of magnitude. Since $\Omega_{\Lambda}$ scales
differently in time with $\Omega_{dark}$ and $\Omega_B$,
the similarity between $\Omega_{\Lambda}$ and $\Omega_{matter} =
\Omega_{dark} + \Omega_B$ might simply be a coincidence.
However,
the similarity in magnitude of the ordinary and dark matter
densities:
\begin{equation}
\Omega_{B}/\Omega_{dark} \approx 0.20,
\end{equation}
is expected to be constant in time
until a very early epoch.
This means that the amount of dark matter produced
in the early universe is of the same order of magnitude
as the ordinary matter, despite their apparent
disparate properties.

The similarity in the abundances of
ordinary and dark matter   
hints at an underlying similarity between the microscopic properties
of the elementary particles comprising the ordinary matter and the dark matter.
Clearly, the standard exotic weakly interacting dark matter
scenarios offer no hope in explaining this cosmic
coincidence because these particles have completely
different properties (different masses and interactions)
from the ordinary baryons. A priori, a dark matter/ordinary
matter ratio of, say, $10^6$ would appear to be equally likely in 
these scenarios.
However if dark matter is identified with mirror baryons, then it
seems to be possible to explain the similarity of $\Omega_{dark}$
and $\Omega_B$ because the microscopic properties of the
mirror particles mirror those of the ordinary particles.
In fact, $\Omega_{dark} = \Omega_B$ would occur 
if the initial conditions of the universe were also mirror symmetric
and no macroscopic asymmetry (such as a temperature difference) 
was produced during the early evolution of the universe.
However, the success of standard big bang nucleosynthesis
(BBN)\cite{sarkar}
does suggest
that $T'$ was somewhat less than $T$ during the BBN epoch,
\begin{eqnarray}
T'/T \stackrel{<}{\sim} 0.6 \ \ {\rm at} \ \ T \sim 1\ MeV,
\label{4}
\end{eqnarray}
in order for the expansion rate of the universe to have been within
an acceptable range\footnote{
If $T' = T$ then the energy density at the BBN epoch 
would be double the standard value -- significantly
increasing the expansion rate of the Universe at that time. 
The equilibration of the three mirror neutrinos,
mirror electron/positron and mirror photons is equivalent to
an extra $\delta N_{\nu} \simeq 6.14$ neutrino species. More generally,
$\delta N_{\nu} = 6.14 (T'/T)^4$, so that demanding $\delta N_{\nu} < 1$
would imply $T' < 0.64T$.}.
If the temperatures are different, then this means
that either the initial conditions of the universe
were asymmetric or that the asymmetry was
induced during the early evolution of the universe.
Actually, within the inflation paradigm it is quite
easy to generate the required temperature asymmetry\cite{inflation}.
In particular, it is natural to have an
`ordinary inflaton' coupling to ordinary matter,
and a `mirror inflaton' coupling to mirror matter.
If inflation is triggered by some random fluctuation, then
it can occur in the two sectors at different times, leading
to $T \neq T'$ after reheating in the two sectors.
In such a scenario, one expects the baryon number and 
mirror baryon number to be unequal (since baryogenesis
or leptogenesis depends on the temperature and expansion rate).

Provided that the temperatures
of the two sectors are not too different  
this might explain the fact that $\Omega_B$ is within an
order of magnitude of $\Omega_{dark}$. 
Clearly, the details will depend on the precise model for
baryogenesis used by nature, which is of course not known
(see the first paper of Ref.\cite{comelli} for a couple of examples).
Even if there is a large hierarchy in temperatures for the
two sectors, similar abundances of ordinary and mirror
particles can be achieved if there are interactions which
can transfer lepton or baryon asymmetry between the
two sectors\cite{fv}. The simplest such interaction is given
by the dimension 5 operator:
\begin{eqnarray}
{\cal L} = {1 \over M_{N}}\bar \ell_L \phi^c \ell'_R \phi' + H.c.
\end{eqnarray}
where $\ell_L$ is a left-handed ordinary lepton doublet, $\ell'_R$
is its mirror partner, $\phi$ is the ordinary Higgs doublet and
$\phi'$ is its mirror partner ($H.c. =$ Hermitian conjugate). 
With such operators and some plausible assumptions
about the physics governing the early evolution of the Universe,
it is even possible\cite{fv} to {\it quantitatively} explain the inferred
non-baryonic dark matter proportion,
$\Omega_B/\Omega_{dark} = 0.20\pm 0.02$.

We will not discuss much more about the very
early Universe (the era prior to BBN). These were the cosmic dark ages 
where much is speculated but little is known. 
A more recent development (in the history of the Universe)
was the formation of large scale structure, which
is also the next topic.

\subsection{Large Scale Structure formation}

We know from measurements of the cosmic microwave background that the
Universe was extraordinarily homogeneous in the past.
At the present epoch, however, the Universe is no longer particularly
homogeneous: it contains galaxies, clusters of galaxies, superclusters
etc. This large scale structure
is believed to arise from
small primordial inhomogeneities that grow via gravitational
instability. However ordinary baryonic density perturbations cannot 
begin to grow until photon decoupling occurs at a temperature
of around $T_{dec} \approx 0.25$ eV, corresponding to a red shift
of $z_{dec} \approx 1100$.
[Prior to photon decoupling, the radiation pressure
prevents the growth of perturbations].
But this is too late: perturbations which have amplitude 
of order $\delta \sim 10^{-5}$ 
(as inferred from the anisotropies of the cosmic microwave background)
do not have enough time to grow into galaxies, where $\delta \sim 10^2$.
This suggests that inhomogeneities begin to grow prior to
photon decoupling. This is one role that non-baryonic
dark matter is expected to fill: it should be weakly coupled
to the ordinary particles in the plasma so that density
perturbation growth can begin prior to photon decoupling.

Mirror particles are weakly coupled to the ordinary ones. 
However, mirror baryonic density
perturbations can only begin to grow after mirror photon
decoupling occurs (roughly when $T'_{dec} \sim 0.25$ eV).
The key point is that if $T' < T$ [which we infer from BBN, see
Eq.(\ref{4})] then mirror photon decoupling necessarily
occurs {\it earlier} than ordinary photon decoupling.
Thus, we expect that mirror baryonic structure formation should
begin earlier than ordinary baryonic structure. According
to Refs.\cite{comelli,rays}, they find that for $T' \stackrel{<}{\sim} 0.2T$,
large scale structure formation with
mirror matter-type dark matter closely resembles the standard cold dark matter 
scenario (but
with some intriguing differences) and
would provide a successful framework to 
understand the large scale structure of the
Universe\cite{comelli,rays,comelli2}.

A consequence of $T' < T$, required for successful big bang
nucleosynthesis and large scale structure formation, is that
mirror BBN occurs earlier than ordinary
BBN. This will mean that the proportion of mirror helium ($He'$) to
mirror hydrogen ($H'$) synthesised in the early Universe will
be different to their ordinary matter
counterparts. In fact, since the expansion rate of
the Universe is faster at earlier times
the mirror neutron/mirror proton ratio should be closer to unity c.f. ordinary
BBN. This means that the $He'/H'$ ratio 
is expected to be significantly greater than the corresponding 
ordinary $He/H$ ratio\cite{comelli}. 
This chemical imbalance between
the ordinary and mirror sectors will no doubt 
have important effects. 
For example,
the initial mass function for mirror stars can be quite different
than for ordinary stars.
Ultimately, this chemical imbalance might even
be responsible for the different distribution of
ordinary and mirror matter in galaxies, as we will
now discuss.

\vskip 0.3cm

\subsection{Asymptotically flat rotation curves and 
the radiative cooling problem}

\vskip 0.3cm

We have briefly mentioned cosmological evidence for
non-baryonic dark matter in sections 2.1 and 2.2.
There is also strong astrophysical evidence for a large amount 
of dark matter in galaxies 
and galaxy clusters. Asymptotically flat rotation curves  
in spiral galaxies, illustrated in Figure 1, imply that 
there must exist invisible `halos' in
galaxies such as our own Milky Way.
These 
halos are, roughly, spherical distributions of invisible matter
which dominate the mass of the galaxy. For example,
the mass of the invisible halo of the Milky Way galaxy is
estimated to be $\sim 6 \times 10^{11} M_{\odot}$, which
is about an order of magnitude more than the estimated mass of the 
galactic disk component\cite{book}.

\vskip 0.2cm
\centerline{\epsfig{file=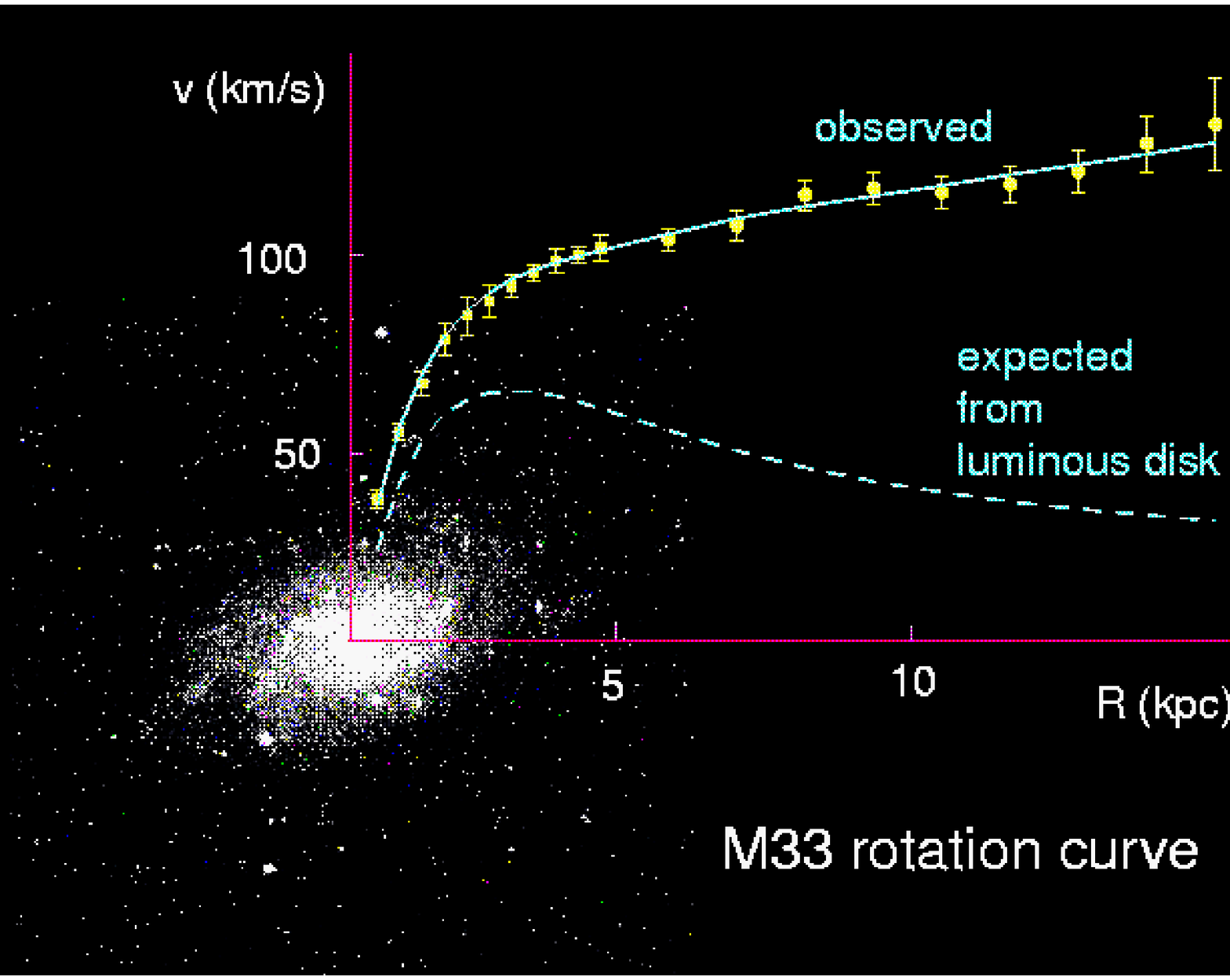,width=9.2cm}} \vskip 0.8cm
\noindent {\small Figure 1: The observed rotation curve for the
the spiral galaxy M33 superimposed on its optical image. 
[Figure from Ref.\cite{fig}].}
\vskip 1.2cm

There is strong
evidence that this galactic halo dark matter must be something exotic;
ordinary baryons simply cannot account for it\cite{freeze}.
Taking the case of white dwarfs as an example, these would be dim
enough to escape detection (unless they are very young).
However, in the collapse process where they are formed the outer layers
of the star are ejected into space.
This ejected material is rich in heavy elements such as oxygen
and nitrogen.
If such material were present in the halo it would have
been revealed from characteristic absorption/emission lines.
Alternatively, if the ejected material were to collapse onto the
galactic disk due to collisional processes, its 
estimated abundance would be greater than the entire mass of the disk.
Thus, old (ordinary) white dwarfs cannot 
provide a consistent picture for halo dark matter.
All other conventional candidates for galactic dark matter run
into similar severe difficulties.

Obviously a (roughly) spherical halo containing mirror stars, mirror planets,
mirror dust and mirror gas would be much less 
problematic since any absorption/emission lines would be
absent\footnote{
Technically, ordinary photon absorption/emission lines would still be there 
due to the effect of photon-mirror photon
kinetic mixing. However the intensity would be reduced by a factor
of $\epsilon^2$ (and $\epsilon^2 \sim 10^{-17}$ given the fit of the DAMA/NaI 
experiment, see later discussion in section 2.5). }.
Of course, there is still the important problem of explaining the 
roughly spherical mirror matter distribution in the galaxy,
with ordinary matter collapsed onto the disk.
A priori this is possible: although ordinary and mirror
matter have identical microscopic interactions, {\it there
is no macroscopic mirror symmetry}. Recall this macroscopic asymmetry
is necessary to explain a) different abundance of
ordinary and mirror matter in the Universe ($\Omega_{dark} \neq 
\Omega_B$, but $\Omega_{dark} \approx 5\Omega_B$) and
b) the different temperatures of the ordinary and mirror
sector (in the early Universe) required by successful BBN
and large scale structure (as discussed earlier).
Because of this macroscopic asymmetry, the evolution of the ordinary
and mirror sectors can be significantly different.

Assuming that the halo is dominated by a mirror gas component which
is approximately spherical and isothermal, its distribution can be
obtained from
the condition of hydrostatic equilibrium\footnote{
The material in this subsection follows Ref.\cite{fvnew}.}
: 
\begin{eqnarray}
{dP \over dr} = -\rho g(r)
\label{pr}
\end{eqnarray}
where $P$ is the pressure and $g(r)$ is the local acceleration at a radius $r$.
For a dilute gas, the pressure
is related to the mass density, $\rho$, via $P = \rho T/(\mu M_p)$, 
where $\mu M_p$ is the average mass
of the particles in the gas ($M_p$ is the proton mass). 
The local acceleration can be simply expressed in terms of the
energy density, via:
\begin{eqnarray}
g(r) = {G \over r^2}\int^r_{0} \rho 4\pi r'^2 dr' \ ,
\label{fin}
\end{eqnarray}
where $G$ is Newton's constant.

The solution of Eq.(\ref{pr},\ref{fin}) is
$\rho \propto 1/r^2$:
\begin{eqnarray}
\rho &=& {\lambda \over r^2}\ ,
\nonumber \\
T &=& G\lambda 2\pi \mu M_p
\ .
\end{eqnarray}
The rotational velocity at a radial location $R_0$,
$v_{rot}(R_0)$, can be obtained from $v_{rot}^2/R_0 = g(R_0)$
which implies:
\begin{eqnarray}
v^2_{rot} (R_0) &=& {G \over R_0}\int^{R_0}_0 \rho 4\pi r^2 dr
\nonumber \\
&=& G\lambda 4\pi .
\end{eqnarray}
Clearly, $\rho = \lambda/r^2$, implied by a spherically symmetric
self gravitating gas in hydrostatic equilibrium, 
gives the required asymptotically flat rotation
curve (a well-known result). Furthermore, from the above 
equation, $\lambda = v_{rot}^2/(4\pi G)$, which 
means that we can express $\rho$ and $T$ in terms
of $v_{rot}$:
\begin{eqnarray}
\rho &=& {v^2_{rot}
\over 4\pi G}{1 \over r^2}
\approx  0.3 \left( {v_{rot} \over
220 \ {\rm km/s}}\right)^2 \left({10 \ {\rm kpc} \over
r}\right)^2 \ {\rm GeV/cm^{3}}
\nonumber \\
T &=& {\mu M_p  v_{rot}^2 \over 2 }
\approx 300 \left( {\mu M_p \over 1 \ {\rm GeV}}\right)\left( {v_{rot}
\over 220 \ {\rm km/s}}\right)^2 \ {\rm eV}
\ .
\label{wed}
\end{eqnarray}
Henceforth we focus on the Milky Way galaxy, for which $v_{rot} \approx 
220$ km/s.

Since $T$ is much greater than the ionization energy for the light
mirror elements ($H', He'$), these elements should
be ionized. It follows that bremsstrahlung and other processes
will radiate off energy at a rate per unit volume of\cite{cdy}:
\begin{eqnarray}
r_{cool} = n_{e'}^2 \Lambda
\end{eqnarray}
where $n_{e'}$ is the (free) mirror electron number density and 
$\Lambda$ is a calculable function (which depends on cross section,
temperature, composition etc).
For a temperature of $T \sim 300$ eV, $\Lambda \sim 
10^{-23}\ {\rm erg\ cm^3 \ s^{-1}}$ (see Ref.\cite{cdy} for
details).

Note that $n_{e'} = 2n_{He'} \simeq 2\rho/M_{He'}$
(for $He'$ mass dominated halo),
which implies
[using Eq.(\ref{wed})]
\begin{eqnarray}
n_{e'} \approx 10^{-1} \ \left( {10 \ {\rm kpc} \over r}\right)^2
\ {\rm cm^{-3} .}
\end{eqnarray}
Because $n_{e'} \propto 1/r^2$,
the total halo luminosity,
\begin{eqnarray}
L_{halo} = \int_{r_{min}} n_{e'}^2 \Lambda 4\pi r^2 dr
\end{eqnarray}
is divergent as $r_{min} \to 0$.
However, the inner region of the galaxy
should contain a high density of mirror dust, stars, supernova,
blackholes etc which will make things very complicated.
Energy sources (such as supernova) can heat
the inner region. This effect, as well as the effect
of mirror dust particles (which can potentially make
the inner region opaque to mirror radiation) could potentially 
lead to an increasing temperature towards the galactic centre --
breaking the isothermal approximation.
This would be consistent with observations which
imply that the rotation curves in spiral galaxies
fall in the inner region (as shown in the example of figure 1), suggesting
that the mass density is not increasing, but roughly 
constant there\cite{salucci}.
In other words, the observations themselves imply that the halo
density appears to be ``heated up"\cite{salucci2}, inexplicable
in the standard cold dark matter scenario, but possible
for mirror matter-type dark matter.

In view of the above discussion, 
we introduce a phenomenological cutoff, $R_1$, and
consider only the energy produced at $r > R_1$.
In this case, the energy radiated from the halo is
roughly
\begin{eqnarray}
L_{halo} &=& \int_{R_1}^{100 {\rm kpc}} n_{e'}^2 \Lambda 4\pi r^2 dr
\nonumber \\
&\sim &
\left({3 \ {\rm kpc} \over R_1}
\right)
10^{44}
\ \ {\rm erg/s}.
\end{eqnarray}
The above calculation assumes that the halo contains only
a gas component. From general considerations, as well
as specific evidence from microlensing studies (as will
be discussed in the following subsection),
a significant component of the halo will be in the form of compact
mirror objects: mirror stars, planets etc.
Furthermore compact mirror objects
can potentially dominate the mass in the inner regions
of the galaxy -- which would alleviate the cooling problem
to some extent. 
Still, a heat source of order $10^{43}-10^{44}$ erg/s
seems to be required to compensate for the energy lost due to 
radiative cooling. 

Supernova offer promising candidate heat sources. An
obvious possibility is that mirror supernova can heat
the halo; during an explosion the outer layers of the
star are ejected into interstellar space, with energy
of order $10^{51}$ erg per explosion.
In order to achieve a rate of around $10^{43}$ erg/s
would require a mirror supernova rate in our galaxy of
around one every few years, which is about an order of
magnitude greater than the rate of ordinary supernova.
Presumably this is possible given the uncertainties
in the mirror sector.
For example, as discussed at the end of the
previous subsection, it is possible that the 
initial mass function for mirror stars
is very different to ordinary stars because of the  different chemical
composition (different light element ratios etc). This
would mean that the rate of ordinary supernova could be quite different
to mirror supernova. 

A more subtle, but equally promising possible
energy source might come from 
ordinary supernova
explosions. Due to the effects
of photon-mirror photon kinetic mixing, Eq.(\ref{km}), in the core of
the supernova a significant fraction, $f'$, of an ordinary
supernova's total energy, $E_{SN}$, can be converted into mirror
photons and mirror electrons/positrons, which can
provide a significant heat source for the halo.
The amount of energy going into the halo from ordinary
supernova explosions, is
of order\cite{fs}:\footnote{
Mirror photons will not be observable to ordinary matter
observers. However, mirror supernova explosions should
produce a significant burst of ordinary $\gamma, e^{\pm}$
particles
which are potentially observable. In fact, 
they may have already been observed in the form of
Gamma Ray Bursts
and positron annihilation radiation from the galactic bulge\cite{fs} 
(this will be briefly reviewed in section 3.1).  }
\begin{eqnarray}
E_{in} &=& f' E_{SN} R_{SN}
\nonumber \\
&=& \left( {f' \over 0.1}\right) \left( {E_{SN} \over 3\times 10^{53} \
{\rm erg}}
\right)\left( {R_{SN} \over 0.01 \ {\rm yr^{-1}}}\right)\ 10^{43}\ {\rm
erg/s.}
\end{eqnarray}
Evidently, ordinary supernova's can potentially supply about
the right amount of energy to replace the energy lost in
radiative cooling, if ordinary supernova's occur at a rate, $R_{SN}$,
of order once per hundred years and of order $10\%$ of the
supernova's energy is converted into mirror $e^{\pm},\gamma'$.

Presumably the heating of the mirror sector
and ordinary sector needs to be different
in order to explain why ordinary matter has
collapsed onto the disk and mirror matter has not. This is
not impossible given the 
lack of macroscopic mirror symmetry, leading to
e.g. asymmetric ordinary and mirror supernova rates.
It seems therefore that asymmetric heating of the ordinary and 
mirror sectors
is feasible and we conclude that mirror matter-type
dark matter is capable of explaining the dark matter halo in spiral
galaxies.

In one sense the existence of a dark halo is more
directly explained within the standard WIMP paradigm.
WIMPs being collisionless are non-dissipative and could not
collapse onto the disk. 
However this cure has serious side-effects which are potentially
fatal.  WIMPs being collisionless
particles are relatively simple, macroscopically, and their
distribution can be predicted. The result is a dark 
matter density profile that goes like\cite{fin1} $\rho \propto
1/r^{\gamma}$, with $\gamma \sim 1.5$.
This prediction is in clear disagreement with the observations
(see e.g. ref.\cite{salucci}).
In other words, the standard WIMP paradigm can simply explain the
existence of dark matter in galaxies, but fails to explain the detailed 
distribution of dark matter within the halo. 
This `fact' seems to support the idea that the dark matter
is, macroscopically, more complicated than collisionless
WIMPs.
More evidence for the complexity of the dark halo
is provided by microlensing surveys of stars in nearby galaxies,
which brings us to the next topic.

\subsection{MACHOs}

If non-baryonic dark matter is identified with mirror matter, 
then a substantial fraction of the non-baryonic dark matter
should be in the form of
compact bodies such as mirror stars. This leads naturally
to an explanation\cite{machoexp}
for the mysterious Massive Astrophysical Compact Halo Objects (or MACHO's)
discovered by the MACHO collaboration.

The MACHO collaboration\cite{macho} has been studying the nature of halo dark 
matter with the gravitational microlensing technique\cite{boh},
using source stars in the Large Magellanic Cloud.
This Australian-American experiment has collected 5.7 years
of data and provided statistically strong evidence for
dark matter in the form of invisible star sized objects which
is what you would expect if there is a significant
amount of mirror matter in our galaxy.
The MACHO collaboration\cite{macho} 
has done a maximum likelihood analysis which implies
a MACHO halo fraction of $f \sim 0.2$ for a typical halo model
with a $95\%$ confidence interval of 
\begin{eqnarray}
0.08 < f < 0.50.
\end{eqnarray}
Their most likely MACHO mass is between $0.15 M_{\odot}$ and
$0.9M_{\odot}$ depending on the halo model.
On the other hand,
the EROS team\cite{erosm} studying microlensing
towards the Small Magellanic Cloud did not find evidence for 
a significant population of compact halo objects. This yielded
a constraint which was however consistent with the
$f \sim 0.2$ halo mass fraction
reported by the MACHO collaboration.
More recently,
a new survey\cite{new} has begun examining stars across the face of M31.
They find significant evidence for a population of halo
microlensing dark matter objects, inferring 
a halo mass fraction of $f = 0.29^{+0.30}_{-0.13}$. This
result is consistent with the positive results of the
MACHO collaboration and provides important independent confirmation
of their positive signal.
Furthermore,
they find significant evidence for an asymmetry in the
distribution of microlensing events across the face of M31,
which is expected if their events are correctly interpreted as 
a large population of invisible massive compact halo objects.

It is important to realize that the 
inferred MACHO halo fraction, $f \sim 0.2$,
is consistent with a mirror matter halo;
the entire halo need not be in the form of mirror stars.
Mirror gas and dust would also be expected as they are a necessary
consequence of stellar evolution and can significantly populate
the halo.

%\newpage

\subsection{Implications of mirror matter-type dark matter for
the DAMA/NaI experiment}

As we have just seen, the
results of the microlensing surveys
suggest a MACHO halo fraction of $f \sim 0.2$.
Within the mirror matter theory, this gets a natural interpretation
in terms of mirror stars, mirror white dwarfs etc. 
The remaining fraction, $1-f \sim 0.8$
should presumably be in the form of mirror gas.
Assuming a roughly spherical and isothermal
distribution for this ionized gas, it would have a 
mass density (at our location, $r \sim 10$ kpc) and temperature
[using Eq.(\ref{wed}]:
\begin{eqnarray}
\rho &\approx & 0.3\ {\rm GeV/cm^3}
\nonumber \\
T & \approx & {\mu M_p v_{rot}^2 \over 2} \sim 300\ {\rm eV}.
\label{wed2}
\end{eqnarray}
Considering a particular chemical element, $A'$ (e.g. $A' =
H', He', O'$ etc), the
velocity distribution 
for these halo mirror particles is then:
\begin{eqnarray}
f_{A'} (v) &=& exp[-{1 \over 2}M_{A'} v^2/T]
\nonumber \\
&\equiv &
exp[-v^2/v_0^2]\ ,
\label{fr5}
\end{eqnarray}
where $v_0^2 \equiv 2T/M_{A'}$. Using Eq.(\ref{wed2}), we have:
\begin{eqnarray}
{v_0^2 (A') \over v_{rot}^2} = {\mu M_p \over M_{A'}}
\ .
\label{z3}
\end{eqnarray}
Recall, $\mu M_p$ is the mean mass of the particles
comprising the mirror (gas) component of the halo 
and $v_{rot} \simeq 220$ km/s is the local rotational
velocity. Evidently the 
characteristic velocity,
$v_0 (A')$,
for a particular halo component element, $A'$,
depends on the chemical composition of the halo
(through the dependence on $\mu M_p$).
Mirror BBN will generate $H', He'$ and, possibly, heavier
mirror elements as well, quite unlike the ordinary matter
case
\footnote{
The abundance of ordinary heavy elements produced during 
BBN is negligible. This is because the number
density is too low for three-body processes, such
as the triple alpha process, to occur at a significant rate.
The situation in the mirror sector could be quite
different.  Because $\Omega_{B'} \approx
5\Omega_{B}$ and also if
$T' \ll T$, then the number density of mirror
nucleons present during the mirror BBN epoch can
be several orders of magnitude greater than the number
density of ordinary nucleons at the time of ordinary
BBN.  The greater
mirror nucleon number density can dramatically increase
the rate of three-body processes such as
the triple alpha process.  Thus, it seems to be an
interesting possibility that a significant abundance
of heavy mirror elements (such as $C', \ O', \ Ne', \ Si'$)
could be generated in the early Universe.}.
Heavy mirror elements can also be generated 
in mirror stars. 
In any case, we consider
two representative possibilities: first that the mass of the halo
is dominated by $He'$ and the second is that
the halo is dominated by $H'$. 
The mean mass of the particles in the halo are then
(taking into
account that the light halo mirror atoms should be fully ionized):
\begin{eqnarray}
\mu M_p &\simeq & 1.3 \ {\rm GeV} 
\ \ {\rm for \ He' \ dominated \ halo,}
\nonumber \\
\mu M_p &\simeq & 0.5 \ {\rm GeV} 
\ \ {\rm for \ H' \ dominated \ halo.}
\end{eqnarray}
The $v_0$ values can
then be easily obtained from Eq.(\ref{z3}):
\begin{eqnarray}
v_0 (A') &=& v_0 (He') \sqrt{{M_{He'}\over M_{A'}}} \approx {220 \over
\sqrt{3}}
\sqrt{{M_{He'}\over M_{A'}}} \ {\rm km/s} \ \  \ \rm{for \ He' \ dominated \
halo}\nonumber \\
v_0 (A') &=& v_0 (H') \sqrt{{M_{H'}\over M_{A'}}} \approx {220
\over \sqrt{2}}
\sqrt{{M_{H'}\over M_{A'}}} \ {\rm km/s} \ \ \ \rm{for \ H' \ dominated \
halo}.
\label{19y}
\end{eqnarray}

It is important to realize that
halo atoms can potentially
be detected in conventional dark matter experiments via 
the nuclear recoil signature\cite{f03}.
The reason is that the photon-mirror photon
kinetic mixing interaction, Eq.(\ref{km}), 
gives the mirror nucleus, with (mirror) atomic number $Z'$, 
a small effective ordinary
electric charge of $\epsilon Z'e$. This means that
ordinary and mirror nuclei can elastically scatter
off each other (essentially Rutherford scattering).
The basic Feynman diagram for this process is given in
the figure 2 (on the following page).

For a mirror atom of mass $M_{A'}$ and (mirror) atomic number 
$Z'$ scattering
on an ordinary target atom of mass $M_{A}$ and atomic number $Z$, 
the differential cross section is given by:
\begin{eqnarray}
{d \sigma \over dE_R} = {\lambda \over E_R^2 v'^2},
\label{cs}
\end{eqnarray}
where $\lambda 
\equiv 2\pi \epsilon^2 \alpha^2 Z^2 Z'^2 F^2_A F^2_{A'}/M_A$ (the
$F_{A,A'}$ are the nuclear form factors\footnote{
Note that the shielding effects of atomic electrons (or 
mirror electrons in the case where the mirror
atom is not fully ionized) can be safely neglected
if the recoil energy of the target nucleus is in the keV
region.}).
In Eq.(\ref{cs}) 
$v'$ is the velocity of the mirror nucleus relative to the Earth
and $E_R$ is the
recoil energy of the ordinary (target) nucleus. 

\vskip 0.7cm
\centerline{\epsfig{file=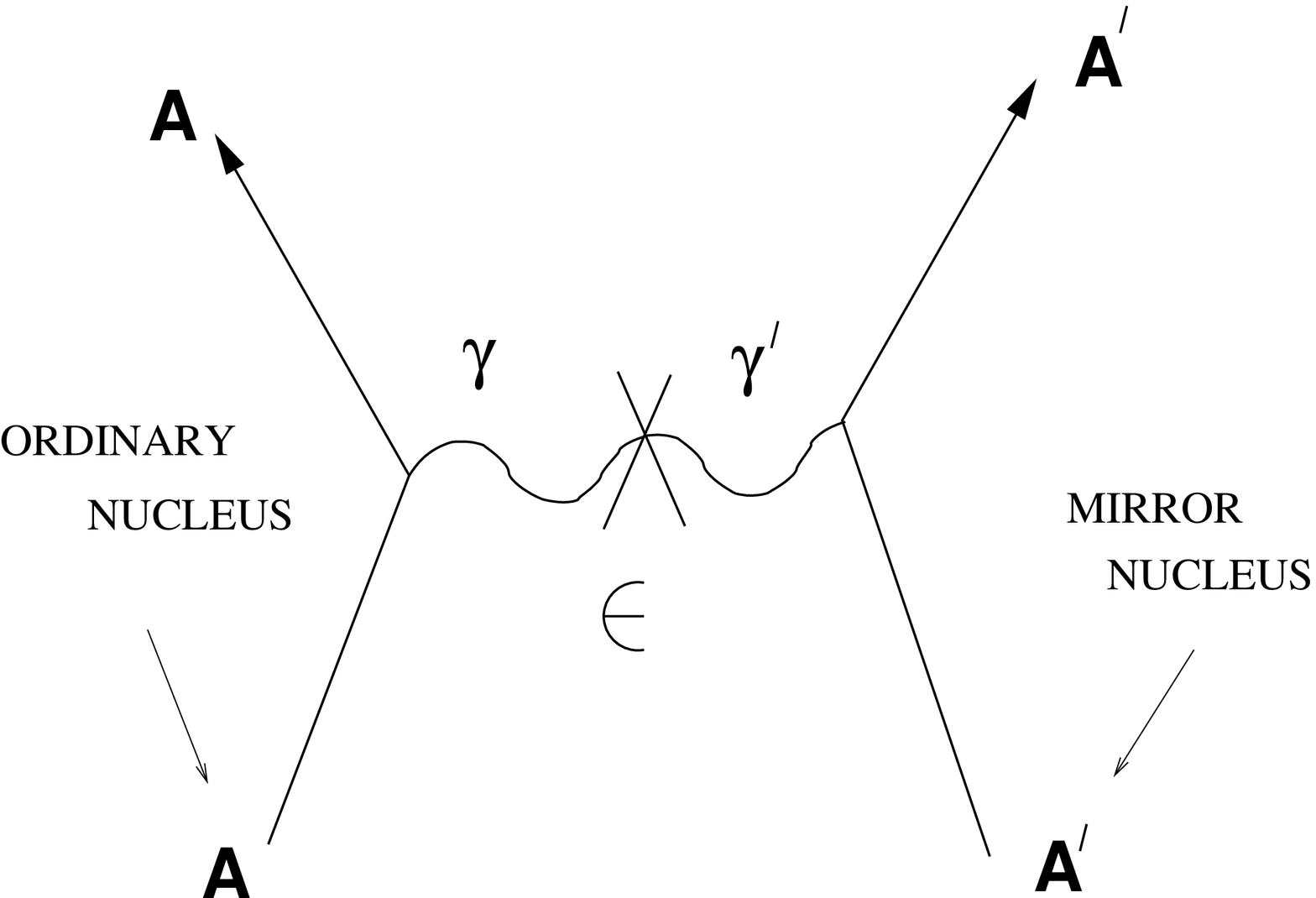,height=4.6cm,width=7.8cm}}
\vskip 0.3cm
\noindent
{\small Figure 2: Ordinary and mirror nuclei can elastically
scatter via the photon-mirror photon kinetic mixing interaction
(indicated by a `cross' in this Feynman diagram).}
\vskip 0.9cm

In dark matter direct detection experiments (such as
DAMA/NaI\cite{dama2}), the measured quantity is
the recoil energy, $E_R$, of the target nucleus. 
The differential interaction rate is
\begin{eqnarray}
{dR \over dE_R} &=&
\sum_{A'} N_T n_{A'} \int^{\infty}_{v'_{min} (E_R)}
{d\sigma \over dE_R} {f(v',v_E) \over k} |v'|
d^3v' \nonumber \\
&=& \sum_{A'} N_T n_{A'}
{\lambda \over E_R^2 } \int^{\infty}_{v'_{min}
(E_R)} {f(v',v_E) \over k|v'|} d^3 v'
\label{55}
\end{eqnarray}
where $N_T$
is the number of target atoms per kg of detector
(for detectors with more than one target element
we must work out the
interaction rate for each element separately and add them up to get the total
interaction rate).
In the above equation
$f(v',v_E)/k$ is the velocity distribution
of the mirror element ($k$ is the normalization factor), 
$A'$, and $v_E$ is the Earth velocity relative to the
galaxy.
Since $v = v' + v_E$ 
is the velocity of the mirror particles relative to the galaxy,
it follows from
Eq.(\ref{fr5}), that 
$f(v',v_E)/k 
= (\pi v_0^2)^{-3/2} exp [ -(v' + v_E)^2/v_0^2]
$.
In Eq.(\ref{55}) the lower velocity limit,
$v'_{min} (E_R)$, is the minimum velocity
for which a mirror atom of mass $M_{A'}$ impacting on
a target atom of mass $M_A$ can produce a recoil
energy of $E_R$ for the target atom. 
This minimum velocity satisfies the
kinematic relation:
\begin{eqnarray}
v'_{min} (E_R) &=& \sqrt{ {(M_A + M_{A'})^2 E_R\over 2M_A M^2_{A'}} } .
\label{v}
\end{eqnarray}
Interestingly, most of the existing dark matter experiments are not very
sensitive to mirror matter-type dark matter because $v'_{min}$
[Eq.(\ref{v})]
turns out to be too high. This is because they either use target
elements which are too heavy (i.e. large $M_A$) and/or have a
$E_R$ threshold which is too high.

The dark matter experiment most sensitive to halo mirror 
matter-type dark matter
is the DAMA/NaI
experiment\cite{dama2}.
The aim of the DAMA/NaI experiment is to measure the
nuclear recoils of $Na, \ I$ atoms due to the interactions
of halo dark matter particles. 
Because of the dependence of the
interaction rate, Eq.(\ref{55}), on $v_E$, 
the interaction rate of halo dark matter with a detector
depends on the Earth's velocity relative to the halo.
Because of the Earth's annual motion, its velocity 
satisfies:
\begin{eqnarray}
v_E (t) & = & v_{\odot} + v_{\oplus} \cos\gamma \cos \omega (t-t_0) 
\nonumber \\
& = &
v_{\odot} + \Delta v_E \cos \omega (t-t_0)
\end{eqnarray}
where $v_{\odot} \approx 230$ km/s is the Sun's velocity with respect
to the galaxy and $v_{\oplus} \simeq 30$ km/s is the Earth's orbital 
velocity around the Sun ($\omega = 2\pi/T$, with $T=1$ year).
The inclination of the Earth's orbital plane relative to the galactic
plane is $\gamma \simeq 60^{o}$, which means that $\Delta v_E \approx
15$ km/s. Thus, the differential
interaction rate in an experiment will thus contain an
annual modulation term:
\begin{eqnarray}
R_i = R_i^0 + R^1_i \cos \omega (t-t_0)
\end{eqnarray}
where
\begin{eqnarray}
R_i^0 &=& {1 \over \Delta E} \int^{E_i + \Delta E}_{E_i}
\left( {dR \over dE_R}\right)_{v_E = v_{\odot}}  dE_R
\nonumber \\
R_i^1 &\simeq & {1 \over \Delta E} \int^{E_i + \Delta E}_{E_i}
{\partial \over \partial v_E} 
\left( {dR \over dE_R}\right)_{v_E = v_{\odot}} \ \Delta v_E dE_R
\ .
\end{eqnarray}
According to the DAMA analysis\cite{dama2}, they indeed find 
an annual modulation
over 7 annual cycles at more than $6\sigma$ C.L. Their data fit 
gives $T = (1.00 \pm 0.01)$ year and 
$t_0 = 144 \pm 22$, consistent with the expected
values. [The expected value for $t_0$ is 152 (2 June), where the
Earth's velocity, $v_E$, reaches
a maximum with respect to the galaxy].  
Their signal occurs in the energy range
$2-6$ keVee\footnote{
The unit keVee corresponds to the detected energy, $\stackrel{\sim}{E}_R$, 
which is related to the actual energy, $E_R$, by $E_R =
\stackrel{\sim}{E}_R/q_{A}$, where $q_A$ is the quenching factor
corresponding to a given target element, $A$. For the DAMA/NaI
experiment, $q_I \simeq 0.09, \ q_{Na} \simeq 0.3$.\cite{dama2}} and
the amplitude of their signal is $R^1 = (0.019 \pm 0.003)$
cpd/kg/keVee [cpd $\equiv$ counts per day].

These are extremely impressive results. The self consistency
of their signal is highly non-trivial: there is simply
no reason why their data should contain a periodic modulation
or why it should peak near June 2. In fact, the 
known systematic 
errors are several orders of magnitude too small to account for
the signal\cite{dama2,sys}.
It therefore seems probable that DAMA has indeed discovered
dark matter.
Interestingly the interpretation of the DAMA/NaI
signal in terms of standard WIMPs appears to be
disfavoured by a number of experiments\cite{cdms,ed,zeplin}, the
most impressive of which is the recent null CDMSII/Ge results\cite{cdms2}. 
However, if
we interpret the DAMA/NaI signal in terms of mirror matter-type
dark matter
then the positive
DAMA/NaI signal and the negative results of the other
experiments can be reconciled\cite{f03,footdama3}.

The DAMA experiment is not particularly sensitive to very
light dark matter particles such as mirror hydrogen and mirror
helium. Impacts of these elements (typically) do not
transfer enough energy to give a signal above the detection
threshold\cite{f03}. 
If stellar nucleosynthesis in the mirror sector is sufficiently
similar to the ordinary sector, then
the next most abundant element should
be mirror oxygen. 
In the analysis of ref.\cite{f03}
the spectrum of heavy
mirror elements were approximated by just two components, $O', Fe'$.
Naturally this is just a crude approximation: in general
there will be a distribution of mirror elements which is
very difficult to theoretically predict (because
of e.g. unknown initial mass function 
for mirror stars etc). 
Of course, it may turn out that approximating the spectrum 
in terms of the two components, $O', Fe'$, 
will be insufficient in the future as more detailed data is
obtained.  Anyway, 
interpreting the DAMA/NaI annual modulation signal in terms
of $O', Fe'$, it was found that\cite{f03}:
\begin{eqnarray}
|\epsilon | \sqrt{ {\xi_{O'} \over 0.10} +
{\xi_{Fe'} \over 0.026}}
\simeq 4.8^{+1.0}_{-1.3} \times 10^{-9}
\label{dama55}
\end{eqnarray}
where the errors denote a 3 sigma allowed range and
$\xi_{A'} \equiv \rho_{A'}/(0.3 \ {\rm GeV/cm^3})$
is the $A'$ proportion (by mass) of the halo dark matter
(at the Earth's location)
\footnote{
The value of $\epsilon$ suggested by the DAMA experiment,
Eq.(\ref{dama55}),
would also have important implications for the orthopositronium
system\cite{fg}. The current experimental situation,
summarized in Ref.\cite{ortho,bader}, implies that
$|\epsilon | \stackrel{<}{\sim} 5\times 10^{-7}$,
which is easily consistent with Eq.(\ref{dama55}).
Importantly, a new orthopositronium experiment
has been proposed\cite{bader} which can potentially
cover much of the $\epsilon$ parameter space suggested by the
DAMA experiment. Such an experiment is
very important -- not just as a check of the
mirror matter explanation -- but also because dark matter experiments
are
sensitive to $\epsilon \sqrt{\xi_{A'}}$ and an independent
measurement of $\epsilon$ would allow the extraction of
$\xi_{A'}$ values.}.
This fit to the data is shown in figure 3. 

\vskip -0.1cm
\centerline{\epsfig{file=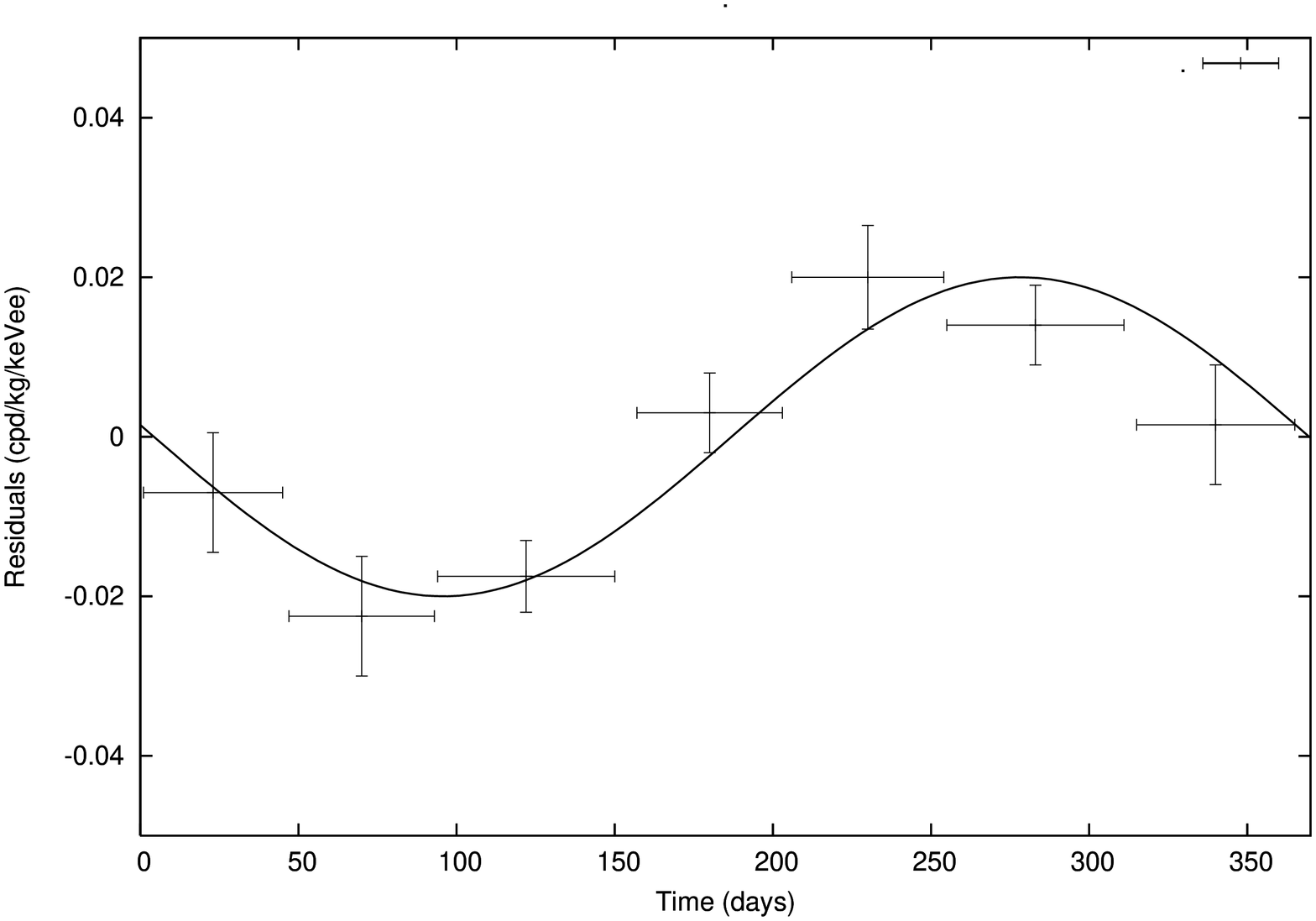,angle=270,width=10.4cm}} 
\vskip 0.2cm
\noindent {\small Figure 3: DAMA/NaI annual modulation signal (taking
data from the second paper of ref.\cite{dama2}) together
with the mirror matter prediction (initial time
is August $7^{th}$).}
\vskip 1.1cm
In Ref.\cite{f03} it was found that a DAMA/NaI
annual modulation signal dominated by an $Fe'$ component,
is experimentally disfavoured for three independent reasons:
a) it predicts a mean differential energy spectrum rate
larger than the measured DAMA/NaI rate
b) potentially leads to a significant diurnal effect (sidereal daily
variation)\footnote{
Currently there is no experimental evidence for any
diurnal variation in the DAMA/NaI data\cite{di}.} and c) should
have been observed in the CDMSI experiment.
Thus it is probable that lighter mirror elements, such as
a mirror oxygen component, dominates the
DAMA annual modulation signal. From Eq.(\ref{dama55})
this means that $\xi_{Fe'} \stackrel{<}{\sim} \xi_{O'}/4$.
Recently, a more stringent limit of $\xi_{Fe'} \stackrel{<}{\sim}
\xi_{O'}/40$ was obtained in Ref.\cite{footdama3} using
the recent CDMSII/Ge result\cite{cdms2}.

If the DAMA signal is dominated by $O'$, then things
depend on only one parameter, $\epsilon \sqrt{\xi_{O'}}$.
This parameter is fixed from the annual modulation
signal, Eq.(\ref{dama55}), which means that the
event rate (due to $O'$ interactions) can
be predicted for other experiments.
It turns out that,
with the exception of one experiment (CRESSTI\cite{cresst}),
all of the other experiments
are not sensitive to $O'$ interactions.
For example, the predicted rate for CDMSII/Ge due to
$O'$
interactions is given in figure 4 (taken from Ref.\cite{footdama3}).

\vskip 0.3cm
\centerline{\epsfig{file=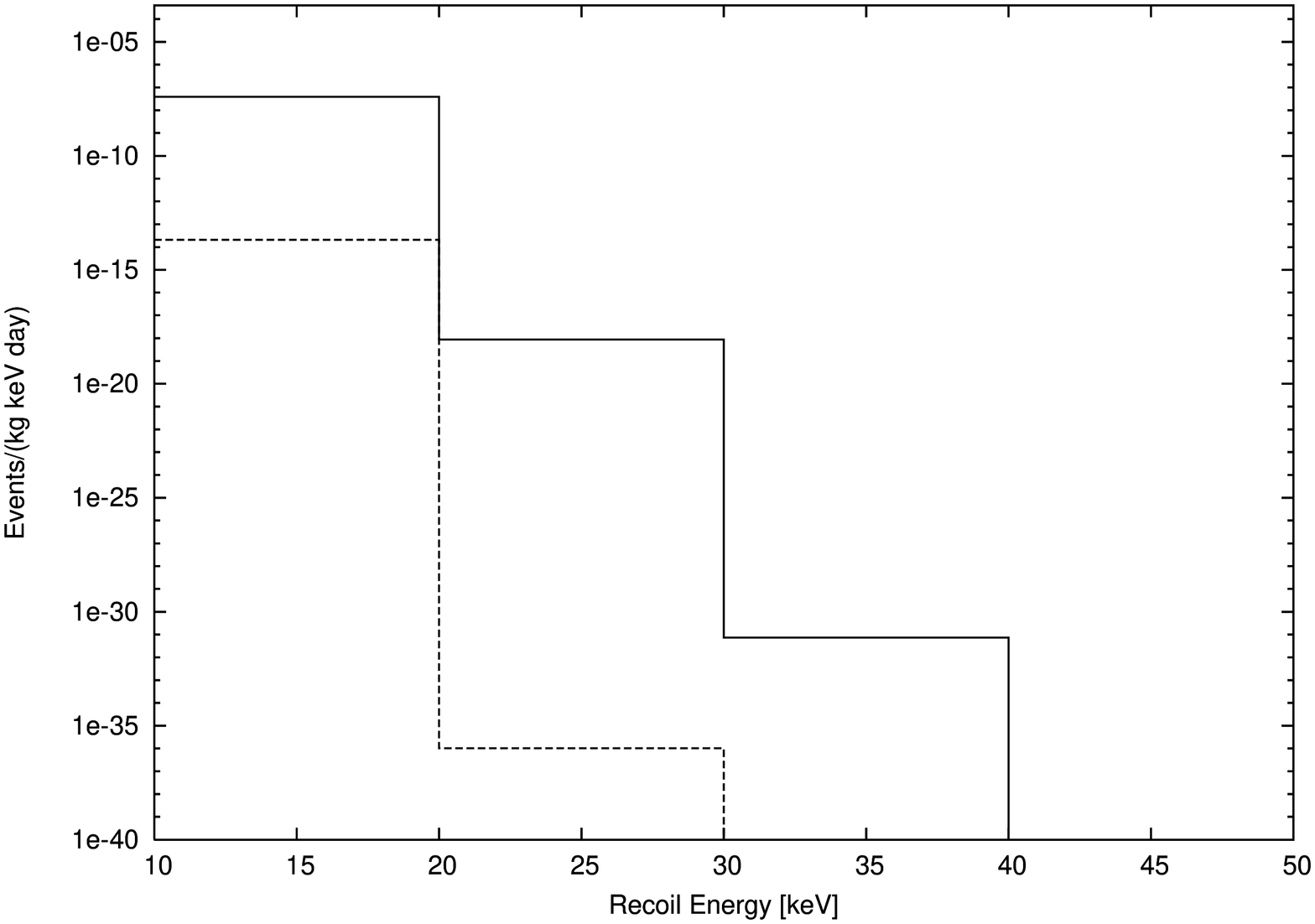,angle=270,width=10.7cm}} \vskip 0.4cm
\noindent {\small Figure 4: 
Predicted differential event rate, $dR/dE_R$, (binned into 10 keV
bins) due to $O'$ dark matter
with $\epsilon \sqrt{\xi_{O'}/0.10} = 4.8\times 10^{-9}$
(DAMA/NaI annual modulation best fit)
for the CDMSII/Ge experiment.
The solid line corresponds to
a standard halo
model with $He'$ dominated halo
while the dashed line assumes
a $H'$ dominated halo.}
\vskip 1.4cm
As the figure shows, the event rate is predicted to be very low.
For the CDMSII/Ge experiment
the predicted event rate is just 1 event per $2.6\times 10^6$ kg-days
for $He'$ dominant halo and 1 event per $5\times 10^{12}$ kg-days if
$H'$ dominates the halo.
Given that CDMSII has only 52.6 kg-day raw exposure in Ge, this implies
a predicted number of events of just $2\times 10^{-5}$ (assuming $He'$
dominant halo)
and even less if $H'$ dominates the mass of the halo.
Clearly this prediction is nicely consistent with the null
result of CDMSII/Ge\footnote{
Although the CDMSII/Ge experiment 
is relatively insensitive to interactions
of halo $O'$, it is much more sensitive to interactions of heavier 
mirror elements such as $Fe'$.
Future CDMS data may well find a positive signal due
to these heavier elements because they should be there at
some level.}.

In the case of standard spin independent WIMPs, the
CDMSII experiment is more sensitive
than the DAMA/NaI experiment. However, as we have discussed
above, this is clearly not the case for $O'$-type
dark matter (with dominant $He'/H'$ component).
The diverse behaviour of the two types of dark matter
candidate has to do with their basic differences:
\begin{itemize}
\item
The mass of $O'$ is only 15 GeV, while standard WIMPs
are typically assumed to have masses which
are greater than $30-45$ GeV (depending on the
model).
\item
For $O'$-type dark matter in an $He'/H'$ dominated halo,
$v_0 (O') \ll 220$ km/s [Eq.(\ref{19y})], while the characteristic
velocity of standard WIMPs are assumed to be approximately 220 km/s.
\item
The differential cross section for
mirror matter-type dark matter is inversely
proportional to the square of the recoil energy, while
that for standard WIMPs is energy independent (excepting the
energy dependence of the form factors).
\end{itemize}
These three key differences mean that experiments with
low threshold energy and light target elements
are much more sensitive (to $O'$-type dark matter) than experiments
with higher threshold energy and/or heavy target
elements. In the case of DAMA/NaI, the event
rate for mirror matter-type dark matter
is dominated by interactions
with the light $Na$ component. The actual threshold energy
of 6.7 keV (for $Na$), implies a threshold impact velocity,
obtained from Eq.(\ref{v}), of
290 km/s for $O'$ impacts. In the case of CDMSII/Ge, the threshold energy
of 10 keV and heavy Ge target gives a threshold impact
velocity of 450 km/s (see Ref.\cite{ortho} for a table
of threshold velocities for the various experiments). Given the
low value of $v_0 (O')$ [Using Eq.(\ref{19y}), 
$v_0 (O') = {110 \over \sqrt{3}}$ km/s
(${55 \over \sqrt{2}}$ km/s) for $He'$
($H'$) dominated halo] the number of $O'$ atoms with impact velocity
above threshold is clearly much lower for CDMSII/Ge compared with
DAMA/NaI (in fact it is exponentially suppressed).
Note that the Edelweiss I/Ge (ref.\cite{ed}) and Zeplin I/Xe
(ref.\cite{zeplin}) experiments
are even less sensitive than CDMSII/Ge because the threshold
impact velocity of those experiments is even higher\cite{ortho}.

There is one experiment, besides DAMA/NaI, which was potentially sensitive
to mirror matter interactions, namely the CRESST I
experiment\cite{cresst}.
That experiment had a target consisting of Sapphire crystals
($Al_2 O_3$), with a low recoil energy threshold
of $0.6$ keV. However, the results of that experiment
turned out to be roughly consistent with the mirror
matter prediction (i.e. with parameters fixed by the
DAMA/NaI annual modulation signal)\cite{f03}, providing tentative
support for the mirror matter interpretation of the DAMA/NaI 
experiment. Unfortunately,
this experiment did not collect
enough data to do
an annual modulation analysis (it has
now been discontinued, replaced by a new experiment, CRESST II,
which
will use a $CaWO_4$ target, and has an expected threshold energy
of 10 keV, which
will be less sensitive than CRESST I, but should still be useful).

\vskip 0.3cm

\section{Unconventional implications of mirror matter-type dark matter
}

\vskip 0.3cm

In section 2 we have examined the conventional cosmological and 
astrophysical implications of mirror matter-type dark matter including 
direct experimental evidence from the DAMA/NaI experiment.
However mirror matter-type dark matter is an unconventional
dark matter candidate with
numerous unconventional implications.
Included among these is the possibility of binary ordinary/mirror
systems, possible manifestations of mirror matter in our
solar system, implications for supernova etc. We now briefly
examine some of these applications.

\subsection{Supernova dynamics, Gamma Ray Bursts and photon-mirror
photon kinetic mixing}

Photon-mirror photon kinetic
mixing, Eq.(\ref{km}), of magnitude $\epsilon \sim 5\times 10^{-9}$
(as suggested by the DAMA annual modulation signal) will lead
to important implications for core collapse supernova -- both ordinary 
and mirror types \footnote{The material in this 
subsection follows Ref.\cite{fs}.}. 
Recall, that in the core of ordinary supernova,
the temperature reaches, $T \sim 30$ MeV, leading to a plasma
of $e^{\pm}, \gamma, \nu_{\alpha}$ ($\alpha = e, \mu, \tau$).
Because of the photon-mirror photon kinetic mixing, mirror
$e'^{\pm}$ 
%(and ultimately $\gamma', \nu'_{\alpha}$ due to 
%mirror electromagnetic and mirror weak interactions)
can also be produced via a variety of processes
(the most obvious being $e^+ + e^- \to e'^+ + e'^-$).
Actually the main production process for mirror particles
in the core of a mirror supernova is expected to be
the plasmon decay process (see
e.g. ref.\cite{raffelt} for a review).
The energy loss rate for production of minicharged
particles has been calculated in Ref.\cite{raffelt}:
\begin{eqnarray}
Q_p = {8\zeta(3) \over 9\pi^3} \epsilon^2 \alpha^2 \left(
\mu_e^2 + {\pi^2T^2 \over 3}\right) T^3 Q_1
\label{Qp}
\end{eqnarray}
where $Q_1$ is a factor of order unity. $Q_p$ is comparable to the
energy loss rate due to neutrino emission for\cite{raffelt}
\begin{eqnarray}
\epsilon  \sim  10^{-9}
\ .
\label{eps}
\end{eqnarray}
Thus, the production of mirror particles 
in the core of ordinary supernova must lead to important effects as
a significant part 
of the emission of ordinary supernova's will be in the form of
$e'^{\pm}, \gamma', \nu'_{\alpha}$. One of these effects is that the 
$e'^{\pm}, \gamma'$ produced in the core will help supernova's
to explode as we will now explain. 

Supernova explosions of
massive stars are believed to be driven  
by the convectivelly supported neutrino-heating
mechanism\cite{collaps}. But refined simulations 
have shown\cite{supern}
that there is insufficient neutrino energy transfer behind the
stalled supernova shock to produce the explosion. This is 
actually a long standing problem in supernova dynamics.
It suggests some missing piece of physics, which might well
be photon-mirror photon kinetic mixing:
the $e'^{\pm}, \gamma'$ produced in the core will interact
and heat the matter behind the shock (adding to the effect
of neutrino-heating) thereby producing the explosion.
For this to be possible we require
that the cross section for
MeV $\gamma'$ (and/or large angle $e'^{\pm}$) scattering
with ordinary electrons (i.e. $\gamma' + e^- 
\to \gamma + e^{-}$ and $e'^{\pm} + e^{-} \to e'^{\pm} + e^{-}$)
to be of roughly the same magnitude as the neutrino nucleon cross
section. The mirror particle cross section is:
%\vskip 0.01cm
\begin{eqnarray}
\sigma \sim \epsilon^2 \pi r_0^2 \sim
10^{-41}\left({\epsilon \over 5\times 10^{-9}}\right)^2 \ {\rm cm}^2
\end{eqnarray}
%\vskip 0.2cm
\noindent
where $r_0 = \alpha/m_e$ is the
classical radius of the electron. The neutrino nucleon cross section 
is
%\vskip 0.01cm
\begin{eqnarray}
\sigma (\bar \nu_e p \to n e^+) &=&
{4G_F^2 E_\nu^2 \over \pi}
\nonumber \\
&\approx &
10^{-41} \left({E_\nu \over 10\ {\rm MeV}}\right)^2 \ {\rm cm}^2
\end{eqnarray}
%\vskip 0.2cm
\noindent
where $E_\nu$ is the energy of the neutrino. 
Evidently the cross sections for the two completely different processes are
indeed comparable!
Importantly, the
energy dependence is different:
compared with neutrino interactions,
the mirror particle interactions with ordinary matter
are larger at lower energies.
It follows that the heating effect of the mirror particle interactions on
the ordinary matter just behind the shock is expected
to be comparable to -- or may even exceed -- the neutrino effect.

A significant portion of the $e'^{\pm}, \gamma'$ will escape
the supernova, however direct detection of these particles seems
to be very difficult for ordinary matter observers. Even if
we cannot directly detect this emission it does not mean
that it is unimportant; as we discussed earlier in section 2.3,
these mirror particles may have an 
important role in heating the galactic halo to
compensate for the energy lost due to radiative cooling.

In the case of a
{\it mirror} type II supernova is also very interesting.
In this case, the core of the mirror
supernova would be a source of ordinary electrons, positrons
and gamma rays -- making such an event easily detectable
for ordinary matter observers. In fact, they may have
already been detected!
Provided that the number of ordinary baryons is sufficiently low 
the $e^+ e^- \gamma$ `fireball' will lead to a gamma ray burst
(GRB)\cite{GRB}.
\footnote{
The idea that GRB's might be connected to mirror supernova was
first suggested by Blinnikov\cite{blin}. 
Blinnikov considered neutrino-mirror neutrino oscillations
(rather than the photon-mirror photon kinetic mixing
interaction) as the mechanism to convert mirror particles into
ordinary particles in the core of a mirror supernova.
Later, it was realized\cite{wong} that neutrino oscillations
were not viable due to matter effects which
strongly suppress neutrino-mirror neutrino oscillations.}
Of course, GRB's have been observed for some time, and their origin
has been a long standing puzzle. It is certainly interesting
that the mirror supernova, with photon-mirror photon kinetic mixing
interaction has roughly
the right characteristics (energy release, time scale, and
potentially small baryon load)
to be identified with the observed gamma ray bursts. 

In addition to being a source of photons, GRB will also eject electrons
and positrons into the interstellar medium. This might explain\cite{fs} the
511 keV photon emission from the galactic bulge. This emission 
was first detected more
than 30 years ago\cite{first} and studied in a number of experiments
culminating in the recent INTEGRAL-SPI measurements\cite{expspi}.

While GRBs and galactic positron emission are certainly rather spectacular
possible manifestations of the mirror world, something even more 
tantalizing would
be the discovery of a mirror world itself.

\subsection{Mirror worlds?}

If mirror matter exists in our galaxy, then
binary systems consisting of ordinary and mirror matter
should also exist. While systems containing approximately
equal amounts of ordinary and mirror matter are very unlikely due
to e.g. differing rates of collapse for
ordinary and mirror matter (due to different 
initial conditions such as chemical composition, temperature
distribution etc),
systems containing predominately ordinary matter 
with a small amount of mirror matter (and vice versa)
should exist. Remarkably, there is interesting evidence for
the existence of such systems coming from extra-solar
planet astronomy.

In 1995, the first planet orbiting another star
was discovered\cite{swiss}. Since that time
the field of extra-solar planet astronomy has
been moving at a rapid pace.
To-date, more than 100 extra-solar planets 
have been discovered orbiting nearby stars\cite{www}. 
They reveal their presence because their gravity tugs
periodically on their parent stars leading to
observable Doppler shifts. Several transiting planets
have been observed 
allowing for an accurate determination of the planet's size and mass
in those systems.
One of the surprising characteristics of the extrasolar planets
is that there are a class of large ($\sim M_{J}, \ J = {\rm Jupiter}$) 
close-in planets with a typical orbital radius of 
$\sim 0.05 \ AU$ (which is about eight times closer than
the orbital radius of Mercury). Ordinary
(gas giant) planets are not expected to form close to stars because the 
high temperatures do not allow them to form.
Theories have been invented where they form far from the
star where the temperature is much lower,
and migrate towards the star\cite{th}. 

A fascinating alternative possibility presents itself in the mirror
world hypothesis. The close-in planets may be mirror worlds
composed predominately of mirror matter\cite{plan}. 
They do not migrate significantly, but actually formed close to 
the star which is not a problem for mirror worlds because
they are not significantly heated by the radiation
from the star. This hypothesis can potentially explain the
opacity of transiting planets because mirror
worlds would accrete ordinary matter from the solar
wind which accumulates in the gravitational potential
well of the mirror world.
It turns out that 
the effective radius, $R_p$ at which the planet becomes
opaque to ordinary radiation depends sensitively
on the mass of the planet, with Ref.\cite{f2} providing
a prediction:
\begin{eqnarray}
R_p \propto \sqrt{{T_s \over M_P} }
\label{1new}
\end{eqnarray}
where $T_s$ is the surface temperature of the planet and
$M_P$ is the mass of the planet.
This was only a rough prediction (especially the
dependence on $T_s$) but a prediction nevertheless.
Heuristically it is very easy to understand: increasing the planet's mass
increases the force of gravity which causes the gas of
ordinary matter to become more tightly bound to the
mirror planet (thereby decreasing the effective size, $R_p$),
while increasing the temperature of the gas increases the volume
that the gas occupies (thereby increasing $R_p$).
Of these two effects we expect that the dependence on $M_P$ should
be the more robust prediction.
Because the size of ordinary gas giant planets (i.e. planets
made mostly of ordinary matter) depends quite weakly on their
mass, the
dependence on $M_P$ -- which is significant according to Eq.(\ref{1new}) --
should allow a decisive test of the mirror planet hypothesis.

There are currently four extrasolar planets for which 
measurements of $R_p$ and
$M_P$ are available: HD209458b\cite{henry}, OGLE-TR-56b\cite{kon}, 
OGLE-TR-113b\cite{bou}
and OGLE-TR-132b\cite{bou}
\
\footnote{The OGLE (= 
Optical Gravitational Lensing Experiment)
transiting planets were identified with
radial Doppler shift measurements of transiting objects discovered
by the Optical Gravitational Lensing Experiment\cite{ogle}.}.
We summarize their properties in the following table:
% SUBJECT: TABLE
\vskip -0.1cm
\begin{table}[h]
\vspace{1mm}
{\centering
\begin{tabular}[t]{|c|c|c|c|}
\hline
{\textbf{Transiting planet}} &
{$R_p$ $[R_J]$} & $M_p$ $[M_J]$ & $T_s$\\
\hline
\hline
{HD209458b} &
{\raggedright $1.43 \pm 0.05$\cite{deeg}}& {\raggedright $0.69 \pm 0.05$
\cite{maz}}
& 1370\ K \\
\hline
{OGLE-TR-56b} &
{\raggedright $1.23 \pm 0.16$}\cite{kon}& {\raggedright $1.45 \pm
0.23$\cite{kon}}
& 1820\ K \\
\hline
{OGLE-TR-113b} &
{\raggedright $1.08 \pm 0.07$\cite{bou}}& {\raggedright $1.35 \pm
0.22$\cite{bou}}
& 1210\ K \\
\hline
{OGLE-TR-132b} &
{\raggedright $1.15^{+0.80}_{-0.13}$\cite{bou}}& {\raggedright $1.01 \pm
0.31$\cite{bou}}
& 1920\ K \\
\hline
\end{tabular}
\par}
%\centering
%\caption
%{The planet radius ($R_p$), mass ($M_p$) and
%effective surface temperature ($T_s$)
%for the four known transiting planets.  }
%\label{table:template}
\vspace{-1mm}
\end{table}
% END TABLE X

\vskip -0.15cm

These measurements (ignoring OGLE-TR-132b because
of its huge uncertainty in $R_p$), together with
the 2001 prediction, Eq.(\ref{1new}), 
are shown in figure 5 (from Ref.\cite{pred}).
The solid line is the prediction, Eq.(\ref{1new}),
where we have used HD209458b to fix the proportionality constant. 
\vskip 0.4cm
\setlength{\unitlength}{0.240900pt}
\ifx\plotpoint\undefined\newsavebox{\plotpoint}\fi
\sbox{\plotpoint}{\rule[-0.200pt]{0.400pt}{0.400pt}}%
\begin{picture}(1500,900)(0,0)
\font\gnuplot=cmr10 at 10pt
\gnuplot
\sbox{\plotpoint}{\rule[-0.200pt]{0.400pt}{0.400pt}}%
\put(181.0,123.0){\rule[-0.200pt]{4.818pt}{0.400pt}}
\put(161,123){\makebox(0,0)[r]{ 0}}
\put(1419.0,123.0){\rule[-0.200pt]{4.818pt}{0.400pt}}
\put(181.0,215.0){\rule[-0.200pt]{4.818pt}{0.400pt}}
\put(161,215){\makebox(0,0)[r]{ 0.2}}
\put(1419.0,215.0){\rule[-0.200pt]{4.818pt}{0.400pt}}
\put(181.0,307.0){\rule[-0.200pt]{4.818pt}{0.400pt}}
\put(161,307){\makebox(0,0)[r]{ 0.4}}
\put(1419.0,307.0){\rule[-0.200pt]{4.818pt}{0.400pt}}
\put(181.0,399.0){\rule[-0.200pt]{4.818pt}{0.400pt}}
\put(161,399){\makebox(0,0)[r]{ 0.6}}
\put(1419.0,399.0){\rule[-0.200pt]{4.818pt}{0.400pt}}
\put(181.0,492.0){\rule[-0.200pt]{4.818pt}{0.400pt}}
\put(161,492){\makebox(0,0)[r]{ 0.8}}
\put(1419.0,492.0){\rule[-0.200pt]{4.818pt}{0.400pt}}
\put(181.0,584.0){\rule[-0.200pt]{4.818pt}{0.400pt}}
\put(161,584){\makebox(0,0)[r]{ 1}}
\put(1419.0,584.0){\rule[-0.200pt]{4.818pt}{0.400pt}}
\put(181.0,676.0){\rule[-0.200pt]{4.818pt}{0.400pt}}
\put(161,676){\makebox(0,0)[r]{ 1.2}}
\put(1419.0,676.0){\rule[-0.200pt]{4.818pt}{0.400pt}}
\put(181.0,768.0){\rule[-0.200pt]{4.818pt}{0.400pt}}
\put(161,768){\makebox(0,0)[r]{ 1.4}}
\put(1419.0,768.0){\rule[-0.200pt]{4.818pt}{0.400pt}}
\put(181.0,860.0){\rule[-0.200pt]{4.818pt}{0.400pt}}
\put(161,860){\makebox(0,0)[r]{ 1.6}}
\put(1419.0,860.0){\rule[-0.200pt]{4.818pt}{0.400pt}}
\put(181.0,123.0){\rule[-0.200pt]{0.400pt}{4.818pt}}
\put(181,82){\makebox(0,0){ 0}}
\put(181.0,840.0){\rule[-0.200pt]{0.400pt}{4.818pt}}
\put(410.0,123.0){\rule[-0.200pt]{0.400pt}{4.818pt}}
\put(410,82){\makebox(0,0){ 0.2}}
\put(410.0,840.0){\rule[-0.200pt]{0.400pt}{4.818pt}}
\put(638.0,123.0){\rule[-0.200pt]{0.400pt}{4.818pt}}
\put(638,82){\makebox(0,0){ 0.4}}
\put(638.0,840.0){\rule[-0.200pt]{0.400pt}{4.818pt}}
\put(867.0,123.0){\rule[-0.200pt]{0.400pt}{4.818pt}}
\put(867,82){\makebox(0,0){ 0.6}}
\put(867.0,840.0){\rule[-0.200pt]{0.400pt}{4.818pt}}
\put(1096.0,123.0){\rule[-0.200pt]{0.400pt}{4.818pt}}
\put(1096,82){\makebox(0,0){ 0.8}}
\put(1096.0,840.0){\rule[-0.200pt]{0.400pt}{4.818pt}}
\put(1325.0,123.0){\rule[-0.200pt]{0.400pt}{4.818pt}}
\put(1325,82){\makebox(0,0){ 1}}
\put(1325.0,840.0){\rule[-0.200pt]{0.400pt}{4.818pt}}
\put(181.0,123.0){\rule[-0.200pt]{303.052pt}{0.400pt}}
\put(1439.0,123.0){\rule[-0.200pt]{0.400pt}{177.543pt}}
\put(181.0,860.0){\rule[-0.200pt]{303.052pt}{0.400pt}}
\put(29,595){\makebox(0,0){$R_p [R_J]$}}
\put(810,21){\makebox(0,0){$\sqrt{T_s/M_P}$}}
\put(181.0,123.0){\rule[-0.200pt]{0.400pt}{177.543pt}}
\put(1325.0,759.0){\rule[-0.200pt]{0.400pt}{11.081pt}}
\put(1315.0,759.0){\rule[-0.200pt]{4.818pt}{0.400pt}}
\put(1315.0,805.0){\rule[-0.200pt]{4.818pt}{0.400pt}}
\put(1090.0,616.0){\rule[-0.200pt]{0.400pt}{35.412pt}}
\put(1080.0,616.0){\rule[-0.200pt]{4.818pt}{0.400pt}}
\put(1080.0,763.0){\rule[-0.200pt]{4.818pt}{0.400pt}}
\put(950.0,588.0){\rule[-0.200pt]{0.400pt}{15.658pt}}
\put(940.0,588.0){\rule[-0.200pt]{4.818pt}{0.400pt}}
\put(940.0,653.0){\rule[-0.200pt]{4.818pt}{0.400pt}}
\put(1285.0,782.0){\rule[-0.200pt]{19.272pt}{0.400pt}}
\put(1285.0,772.0){\rule[-0.200pt]{0.400pt}{4.818pt}}
\put(1365.0,772.0){\rule[-0.200pt]{0.400pt}{4.818pt}}
\put(1033.0,690.0){\rule[-0.200pt]{27.463pt}{0.400pt}}
\put(1033.0,680.0){\rule[-0.200pt]{0.400pt}{4.818pt}}
\put(1147.0,680.0){\rule[-0.200pt]{0.400pt}{4.818pt}}
\put(866.0,620.0){\rule[-0.200pt]{40.230pt}{0.400pt}}
\put(866.0,610.0){\rule[-0.200pt]{0.400pt}{4.818pt}}
%\put(1325,782){\raisebox{-.8pt}{\makebox(0,0){$\Diamond$}}}
%\put(1090,690){\raisebox{-.8pt}{\makebox(0,0){$\Diamond$}}}
%\put(950,620){\raisebox{-.8pt}{\makebox(0,0){$\Diamond$}}}
\put(1033.0,610.0){\rule[-0.200pt]{0.400pt}{4.818pt}}
\put(181.0,123.0){\rule[-0.200pt]{4.818pt}{0.400pt}}
\put(161,123){\makebox(0,0)[r]{ 0}}
\put(1419.0,123.0){\rule[-0.200pt]{4.818pt}{0.400pt}}
\put(181.0,215.0){\rule[-0.200pt]{4.818pt}{0.400pt}}
\put(161,215){\makebox(0,0)[r]{ 0.2}}
\put(1419.0,215.0){\rule[-0.200pt]{4.818pt}{0.400pt}}
\put(181.0,307.0){\rule[-0.200pt]{4.818pt}{0.400pt}}
\put(161,307){\makebox(0,0)[r]{ 0.4}}
\put(1419.0,307.0){\rule[-0.200pt]{4.818pt}{0.400pt}}
\put(181.0,399.0){\rule[-0.200pt]{4.818pt}{0.400pt}}
\put(161,399){\makebox(0,0)[r]{ 0.6}}
\put(1419.0,399.0){\rule[-0.200pt]{4.818pt}{0.400pt}}
\put(181.0,492.0){\rule[-0.200pt]{4.818pt}{0.400pt}}
\put(161,492){\makebox(0,0)[r]{ 0.8}}
\put(1419.0,492.0){\rule[-0.200pt]{4.818pt}{0.400pt}}
\put(181.0,584.0){\rule[-0.200pt]{4.818pt}{0.400pt}}
\put(161,584){\makebox(0,0)[r]{ 1}}
\put(1419.0,584.0){\rule[-0.200pt]{4.818pt}{0.400pt}}
\put(181.0,676.0){\rule[-0.200pt]{4.818pt}{0.400pt}}
\put(161,676){\makebox(0,0)[r]{ 1.2}}
\put(1419.0,676.0){\rule[-0.200pt]{4.818pt}{0.400pt}}
\put(181.0,768.0){\rule[-0.200pt]{4.818pt}{0.400pt}}
\put(161,768){\makebox(0,0)[r]{ 1.4}}
\put(1419.0,768.0){\rule[-0.200pt]{4.818pt}{0.400pt}}
\put(181.0,860.0){\rule[-0.200pt]{4.818pt}{0.400pt}}
\put(161,860){\makebox(0,0)[r]{ 1.6}}
\put(1419.0,860.0){\rule[-0.200pt]{4.818pt}{0.400pt}}
\put(181.0,123.0){\rule[-0.200pt]{0.400pt}{4.818pt}}
\put(181,82){\makebox(0,0){ 0}}
\put(181.0,840.0){\rule[-0.200pt]{0.400pt}{4.818pt}}
\put(410.0,123.0){\rule[-0.200pt]{0.400pt}{4.818pt}}
\put(410,82){\makebox(0,0){ 0.2}}
\put(410.0,840.0){\rule[-0.200pt]{0.400pt}{4.818pt}}
\put(638.0,123.0){\rule[-0.200pt]{0.400pt}{4.818pt}}
\put(638,82){\makebox(0,0){ 0.4}}
\put(638.0,840.0){\rule[-0.200pt]{0.400pt}{4.818pt}}
\put(867.0,123.0){\rule[-0.200pt]{0.400pt}{4.818pt}}
\put(867,82){\makebox(0,0){ 0.6}}
\put(867.0,840.0){\rule[-0.200pt]{0.400pt}{4.818pt}}
\put(1096.0,123.0){\rule[-0.200pt]{0.400pt}{4.818pt}}
\put(1096,82){\makebox(0,0){ 0.8}}
\put(1096.0,840.0){\rule[-0.200pt]{0.400pt}{4.818pt}}
\put(1325.0,123.0){\rule[-0.200pt]{0.400pt}{4.818pt}}
\put(1325,82){\makebox(0,0){ 1}}
\put(1325.0,840.0){\rule[-0.200pt]{0.400pt}{4.818pt}}
\put(181.0,123.0){\rule[-0.200pt]{303.052pt}{0.400pt}}
\put(1439.0,123.0){\rule[-0.200pt]{0.400pt}{177.543pt}}
\put(181.0,860.0){\rule[-0.200pt]{303.052pt}{0.400pt}}
\put(29,595){\makebox(0,0){$R_p [R_J]$}}
\put(810,21){\makebox(0,0){$\sqrt{T_s/M_P}$}}
\put(181.0,123.0){\rule[-0.200pt]{0.400pt}{177.543pt}}
\put(181,123){\usebox{\plotpoint}}
\multiput(181.00,123.59)(0.950,0.485){11}{\rule{0.843pt}{0.117pt}}
\multiput(181.00,122.17)(11.251,7.000){2}{\rule{0.421pt}{0.400pt}}
\multiput(194.00,130.59)(0.758,0.488){13}{\rule{0.700pt}{0.117pt}}
\multiput(194.00,129.17)(10.547,8.000){2}{\rule{0.350pt}{0.400pt}}
\multiput(206.00,138.59)(0.950,0.485){11}{\rule{0.843pt}{0.117pt}}
\multiput(206.00,137.17)(11.251,7.000){2}{\rule{0.421pt}{0.400pt}}
\multiput(219.00,145.59)(0.950,0.485){11}{\rule{0.843pt}{0.117pt}}
\multiput(219.00,144.17)(11.251,7.000){2}{\rule{0.421pt}{0.400pt}}
\multiput(232.00,152.59)(0.824,0.488){13}{\rule{0.750pt}{0.117pt}}
\multiput(232.00,151.17)(11.443,8.000){2}{\rule{0.375pt}{0.400pt}}
\multiput(245.00,160.59)(0.874,0.485){11}{\rule{0.786pt}{0.117pt}}
\multiput(245.00,159.17)(10.369,7.000){2}{\rule{0.393pt}{0.400pt}}
\multiput(257.00,167.59)(0.950,0.485){11}{\rule{0.843pt}{0.117pt}}
\multiput(257.00,166.17)(11.251,7.000){2}{\rule{0.421pt}{0.400pt}}
\multiput(270.00,174.59)(0.824,0.488){13}{\rule{0.750pt}{0.117pt}}
\multiput(270.00,173.17)(11.443,8.000){2}{\rule{0.375pt}{0.400pt}}
\multiput(283.00,182.59)(0.874,0.485){11}{\rule{0.786pt}{0.117pt}}
\multiput(283.00,181.17)(10.369,7.000){2}{\rule{0.393pt}{0.400pt}}
\multiput(295.00,189.59)(0.950,0.485){11}{\rule{0.843pt}{0.117pt}}
\multiput(295.00,188.17)(11.251,7.000){2}{\rule{0.421pt}{0.400pt}}
\multiput(308.00,196.59)(0.824,0.488){13}{\rule{0.750pt}{0.117pt}}
\multiput(308.00,195.17)(11.443,8.000){2}{\rule{0.375pt}{0.400pt}}
\multiput(321.00,204.59)(0.874,0.485){11}{\rule{0.786pt}{0.117pt}}
\multiput(321.00,203.17)(10.369,7.000){2}{\rule{0.393pt}{0.400pt}}
\multiput(333.00,211.59)(0.950,0.485){11}{\rule{0.843pt}{0.117pt}}
\multiput(333.00,210.17)(11.251,7.000){2}{\rule{0.421pt}{0.400pt}}
\multiput(346.00,218.59)(0.950,0.485){11}{\rule{0.843pt}{0.117pt}}
\multiput(346.00,217.17)(11.251,7.000){2}{\rule{0.421pt}{0.400pt}}
\multiput(359.00,225.59)(0.824,0.488){13}{\rule{0.750pt}{0.117pt}}
\multiput(359.00,224.17)(11.443,8.000){2}{\rule{0.375pt}{0.400pt}}
\multiput(372.00,233.59)(0.874,0.485){11}{\rule{0.786pt}{0.117pt}}
\multiput(372.00,232.17)(10.369,7.000){2}{\rule{0.393pt}{0.400pt}}
\multiput(384.00,240.59)(0.950,0.485){11}{\rule{0.843pt}{0.117pt}}
\multiput(384.00,239.17)(11.251,7.000){2}{\rule{0.421pt}{0.400pt}}
\multiput(397.00,247.59)(0.824,0.488){13}{\rule{0.750pt}{0.117pt}}
\multiput(397.00,246.17)(11.443,8.000){2}{\rule{0.375pt}{0.400pt}}
\multiput(410.00,255.59)(0.874,0.485){11}{\rule{0.786pt}{0.117pt}}
\multiput(410.00,254.17)(10.369,7.000){2}{\rule{0.393pt}{0.400pt}}
\multiput(422.00,262.59)(0.950,0.485){11}{\rule{0.843pt}{0.117pt}}
\multiput(422.00,261.17)(11.251,7.000){2}{\rule{0.421pt}{0.400pt}}
\multiput(435.00,269.59)(0.824,0.488){13}{\rule{0.750pt}{0.117pt}}
\multiput(435.00,268.17)(11.443,8.000){2}{\rule{0.375pt}{0.400pt}}
\multiput(448.00,277.59)(0.950,0.485){11}{\rule{0.843pt}{0.117pt}}
\multiput(448.00,276.17)(11.251,7.000){2}{\rule{0.421pt}{0.400pt}}
\multiput(461.00,284.59)(0.874,0.485){11}{\rule{0.786pt}{0.117pt}}
\multiput(461.00,283.17)(10.369,7.000){2}{\rule{0.393pt}{0.400pt}}
\multiput(473.00,291.59)(0.824,0.488){13}{\rule{0.750pt}{0.117pt}}
\multiput(473.00,290.17)(11.443,8.000){2}{\rule{0.375pt}{0.400pt}}
\multiput(486.00,299.59)(0.950,0.485){11}{\rule{0.843pt}{0.117pt}}
\multiput(486.00,298.17)(11.251,7.000){2}{\rule{0.421pt}{0.400pt}}
\multiput(499.00,306.59)(0.874,0.485){11}{\rule{0.786pt}{0.117pt}}
\multiput(499.00,305.17)(10.369,7.000){2}{\rule{0.393pt}{0.400pt}}
\multiput(511.00,313.59)(0.824,0.488){13}{\rule{0.750pt}{0.117pt}}
\multiput(511.00,312.17)(11.443,8.000){2}{\rule{0.375pt}{0.400pt}}
\multiput(524.00,321.59)(0.950,0.485){11}{\rule{0.843pt}{0.117pt}}
\multiput(524.00,320.17)(11.251,7.000){2}{\rule{0.421pt}{0.400pt}}
\multiput(537.00,328.59)(0.950,0.485){11}{\rule{0.843pt}{0.117pt}}
\multiput(537.00,327.17)(11.251,7.000){2}{\rule{0.421pt}{0.400pt}}
\multiput(550.00,335.59)(0.758,0.488){13}{\rule{0.700pt}{0.117pt}}
\multiput(550.00,334.17)(10.547,8.000){2}{\rule{0.350pt}{0.400pt}}
\multiput(562.00,343.59)(0.950,0.485){11}{\rule{0.843pt}{0.117pt}}
\multiput(562.00,342.17)(11.251,7.000){2}{\rule{0.421pt}{0.400pt}}
\multiput(575.00,350.59)(0.950,0.485){11}{\rule{0.843pt}{0.117pt}}
\multiput(575.00,349.17)(11.251,7.000){2}{\rule{0.421pt}{0.400pt}}
\multiput(588.00,357.59)(0.758,0.488){13}{\rule{0.700pt}{0.117pt}}
\multiput(588.00,356.17)(10.547,8.000){2}{\rule{0.350pt}{0.400pt}}
\multiput(600.00,365.59)(0.950,0.485){11}{\rule{0.843pt}{0.117pt}}
\multiput(600.00,364.17)(11.251,7.000){2}{\rule{0.421pt}{0.400pt}}
\multiput(613.00,372.59)(0.950,0.485){11}{\rule{0.843pt}{0.117pt}}
\multiput(613.00,371.17)(11.251,7.000){2}{\rule{0.421pt}{0.400pt}}
\multiput(626.00,379.59)(0.874,0.485){11}{\rule{0.786pt}{0.117pt}}
\multiput(626.00,378.17)(10.369,7.000){2}{\rule{0.393pt}{0.400pt}}
\multiput(638.00,386.59)(0.824,0.488){13}{\rule{0.750pt}{0.117pt}}
\multiput(638.00,385.17)(11.443,8.000){2}{\rule{0.375pt}{0.400pt}}
\multiput(651.00,394.59)(0.950,0.485){11}{\rule{0.843pt}{0.117pt}}
\multiput(651.00,393.17)(11.251,7.000){2}{\rule{0.421pt}{0.400pt}}
\multiput(664.00,401.59)(0.950,0.485){11}{\rule{0.843pt}{0.117pt}}
\multiput(664.00,400.17)(11.251,7.000){2}{\rule{0.421pt}{0.400pt}}
\multiput(677.00,408.59)(0.758,0.488){13}{\rule{0.700pt}{0.117pt}}
\multiput(677.00,407.17)(10.547,8.000){2}{\rule{0.350pt}{0.400pt}}
\multiput(689.00,416.59)(0.950,0.485){11}{\rule{0.843pt}{0.117pt}}
\multiput(689.00,415.17)(11.251,7.000){2}{\rule{0.421pt}{0.400pt}}
\multiput(702.00,423.59)(0.950,0.485){11}{\rule{0.843pt}{0.117pt}}
\multiput(702.00,422.17)(11.251,7.000){2}{\rule{0.421pt}{0.400pt}}
\multiput(715.00,430.59)(0.758,0.488){13}{\rule{0.700pt}{0.117pt}}
\multiput(715.00,429.17)(10.547,8.000){2}{\rule{0.350pt}{0.400pt}}
\multiput(727.00,438.59)(0.950,0.485){11}{\rule{0.843pt}{0.117pt}}
\multiput(727.00,437.17)(11.251,7.000){2}{\rule{0.421pt}{0.400pt}}
\multiput(740.00,445.59)(0.950,0.485){11}{\rule{0.843pt}{0.117pt}}
\multiput(740.00,444.17)(11.251,7.000){2}{\rule{0.421pt}{0.400pt}}
\multiput(753.00,452.59)(0.824,0.488){13}{\rule{0.750pt}{0.117pt}}
\multiput(753.00,451.17)(11.443,8.000){2}{\rule{0.375pt}{0.400pt}}
\multiput(766.00,460.59)(0.874,0.485){11}{\rule{0.786pt}{0.117pt}}
\multiput(766.00,459.17)(10.369,7.000){2}{\rule{0.393pt}{0.400pt}}
\multiput(778.00,467.59)(0.950,0.485){11}{\rule{0.843pt}{0.117pt}}
\multiput(778.00,466.17)(11.251,7.000){2}{\rule{0.421pt}{0.400pt}}
\multiput(791.00,474.59)(0.824,0.488){13}{\rule{0.750pt}{0.117pt}}
\multiput(791.00,473.17)(11.443,8.000){2}{\rule{0.375pt}{0.400pt}}
\multiput(804.00,482.59)(0.874,0.485){11}{\rule{0.786pt}{0.117pt}}
\multiput(804.00,481.17)(10.369,7.000){2}{\rule{0.393pt}{0.400pt}}
\multiput(816.00,489.59)(0.950,0.485){11}{\rule{0.843pt}{0.117pt}}
\multiput(816.00,488.17)(11.251,7.000){2}{\rule{0.421pt}{0.400pt}}
\multiput(829.00,496.59)(0.824,0.488){13}{\rule{0.750pt}{0.117pt}}
\multiput(829.00,495.17)(11.443,8.000){2}{\rule{0.375pt}{0.400pt}}
\multiput(842.00,504.59)(0.874,0.485){11}{\rule{0.786pt}{0.117pt}}
\multiput(842.00,503.17)(10.369,7.000){2}{\rule{0.393pt}{0.400pt}}
\multiput(854.00,511.59)(0.950,0.485){11}{\rule{0.843pt}{0.117pt}}
\multiput(854.00,510.17)(11.251,7.000){2}{\rule{0.421pt}{0.400pt}}
\multiput(867.00,518.59)(0.824,0.488){13}{\rule{0.750pt}{0.117pt}}
\multiput(867.00,517.17)(11.443,8.000){2}{\rule{0.375pt}{0.400pt}}
\multiput(880.00,526.59)(0.950,0.485){11}{\rule{0.843pt}{0.117pt}}
\multiput(880.00,525.17)(11.251,7.000){2}{\rule{0.421pt}{0.400pt}}
\multiput(893.00,533.59)(0.874,0.485){11}{\rule{0.786pt}{0.117pt}}
\multiput(893.00,532.17)(10.369,7.000){2}{\rule{0.393pt}{0.400pt}}
\multiput(905.00,540.59)(0.950,0.485){11}{\rule{0.843pt}{0.117pt}}
\multiput(905.00,539.17)(11.251,7.000){2}{\rule{0.421pt}{0.400pt}}
\multiput(918.00,547.59)(0.824,0.488){13}{\rule{0.750pt}{0.117pt}}
\multiput(918.00,546.17)(11.443,8.000){2}{\rule{0.375pt}{0.400pt}}
\multiput(931.00,555.59)(0.874,0.485){11}{\rule{0.786pt}{0.117pt}}
\multiput(931.00,554.17)(10.369,7.000){2}{\rule{0.393pt}{0.400pt}}
\multiput(943.00,562.59)(0.950,0.485){11}{\rule{0.843pt}{0.117pt}}
\multiput(943.00,561.17)(11.251,7.000){2}{\rule{0.421pt}{0.400pt}}
\multiput(956.00,569.59)(0.824,0.488){13}{\rule{0.750pt}{0.117pt}}
\multiput(956.00,568.17)(11.443,8.000){2}{\rule{0.375pt}{0.400pt}}
\multiput(969.00,577.59)(0.950,0.485){11}{\rule{0.843pt}{0.117pt}}
\multiput(969.00,576.17)(11.251,7.000){2}{\rule{0.421pt}{0.400pt}}
\multiput(982.00,584.59)(0.874,0.485){11}{\rule{0.786pt}{0.117pt}}
\multiput(982.00,583.17)(10.369,7.000){2}{\rule{0.393pt}{0.400pt}}
\multiput(994.00,591.59)(0.824,0.488){13}{\rule{0.750pt}{0.117pt}}
\multiput(994.00,590.17)(11.443,8.000){2}{\rule{0.375pt}{0.400pt}}
\multiput(1007.00,599.59)(0.950,0.485){11}{\rule{0.843pt}{0.117pt}}
\multiput(1007.00,598.17)(11.251,7.000){2}{\rule{0.421pt}{0.400pt}}
\multiput(1020.00,606.59)(0.874,0.485){11}{\rule{0.786pt}{0.117pt}}
\multiput(1020.00,605.17)(10.369,7.000){2}{\rule{0.393pt}{0.400pt}}
\multiput(1032.00,613.59)(0.824,0.488){13}{\rule{0.750pt}{0.117pt}}
\multiput(1032.00,612.17)(11.443,8.000){2}{\rule{0.375pt}{0.400pt}}
\multiput(1045.00,621.59)(0.950,0.485){11}{\rule{0.843pt}{0.117pt}}
\multiput(1045.00,620.17)(11.251,7.000){2}{\rule{0.421pt}{0.400pt}}
\multiput(1058.00,628.59)(0.874,0.485){11}{\rule{0.786pt}{0.117pt}}
\multiput(1058.00,627.17)(10.369,7.000){2}{\rule{0.393pt}{0.400pt}}
\multiput(1070.00,635.59)(0.824,0.488){13}{\rule{0.750pt}{0.117pt}}
\multiput(1070.00,634.17)(11.443,8.000){2}{\rule{0.375pt}{0.400pt}}
\multiput(1083.00,643.59)(0.950,0.485){11}{\rule{0.843pt}{0.117pt}}
\multiput(1083.00,642.17)(11.251,7.000){2}{\rule{0.421pt}{0.400pt}}
\multiput(1096.00,650.59)(0.950,0.485){11}{\rule{0.843pt}{0.117pt}}
\multiput(1096.00,649.17)(11.251,7.000){2}{\rule{0.421pt}{0.400pt}}
\multiput(1109.00,657.59)(0.758,0.488){13}{\rule{0.700pt}{0.117pt}}
\multiput(1109.00,656.17)(10.547,8.000){2}{\rule{0.350pt}{0.400pt}}
\multiput(1121.00,665.59)(0.950,0.485){11}{\rule{0.843pt}{0.117pt}}
\multiput(1121.00,664.17)(11.251,7.000){2}{\rule{0.421pt}{0.400pt}}
\multiput(1134.00,672.59)(0.950,0.485){11}{\rule{0.843pt}{0.117pt}}
\multiput(1134.00,671.17)(11.251,7.000){2}{\rule{0.421pt}{0.400pt}}
\multiput(1147.00,679.59)(0.758,0.488){13}{\rule{0.700pt}{0.117pt}}
\multiput(1147.00,678.17)(10.547,8.000){2}{\rule{0.350pt}{0.400pt}}
\multiput(1159.00,687.59)(0.950,0.485){11}{\rule{0.843pt}{0.117pt}}
\multiput(1159.00,686.17)(11.251,7.000){2}{\rule{0.421pt}{0.400pt}}
\multiput(1172.00,694.59)(0.950,0.485){11}{\rule{0.843pt}{0.117pt}}
\multiput(1172.00,693.17)(11.251,7.000){2}{\rule{0.421pt}{0.400pt}}
\multiput(1185.00,701.59)(0.824,0.488){13}{\rule{0.750pt}{0.117pt}}
\multiput(1185.00,700.17)(11.443,8.000){2}{\rule{0.375pt}{0.400pt}}
\multiput(1198.00,709.59)(0.874,0.485){11}{\rule{0.786pt}{0.117pt}}
\multiput(1198.00,708.17)(10.369,7.000){2}{\rule{0.393pt}{0.400pt}}
\multiput(1210.00,716.59)(0.950,0.485){11}{\rule{0.843pt}{0.117pt}}
\multiput(1210.00,715.17)(11.251,7.000){2}{\rule{0.421pt}{0.400pt}}
\multiput(1223.00,723.59)(0.950,0.485){11}{\rule{0.843pt}{0.117pt}}
\multiput(1223.00,722.17)(11.251,7.000){2}{\rule{0.421pt}{0.400pt}}
\multiput(1236.00,730.59)(0.758,0.488){13}{\rule{0.700pt}{0.117pt}}
\multiput(1236.00,729.17)(10.547,8.000){2}{\rule{0.350pt}{0.400pt}}
\multiput(1248.00,738.59)(0.950,0.485){11}{\rule{0.843pt}{0.117pt}}
\multiput(1248.00,737.17)(11.251,7.000){2}{\rule{0.421pt}{0.400pt}}
\multiput(1261.00,745.59)(0.950,0.485){11}{\rule{0.843pt}{0.117pt}}
\multiput(1261.00,744.17)(11.251,7.000){2}{\rule{0.421pt}{0.400pt}}
\multiput(1274.00,752.59)(0.824,0.488){13}{\rule{0.750pt}{0.117pt}}
\multiput(1274.00,751.17)(11.443,8.000){2}{\rule{0.375pt}{0.400pt}}
\multiput(1287.00,760.59)(0.874,0.485){11}{\rule{0.786pt}{0.117pt}}
\multiput(1287.00,759.17)(10.369,7.000){2}{\rule{0.393pt}{0.400pt}}
\multiput(1299.00,767.59)(0.950,0.485){11}{\rule{0.843pt}{0.117pt}}
\multiput(1299.00,766.17)(11.251,7.000){2}{\rule{0.421pt}{0.400pt}}
\multiput(1312.00,774.59)(0.824,0.488){13}{\rule{0.750pt}{0.117pt}}
\multiput(1312.00,773.17)(11.443,8.000){2}{\rule{0.375pt}{0.400pt}}
\multiput(1325.00,782.59)(0.874,0.485){11}{\rule{0.786pt}{0.117pt}}
\multiput(1325.00,781.17)(10.369,7.000){2}{\rule{0.393pt}{0.400pt}}
\multiput(1337.00,789.59)(0.950,0.485){11}{\rule{0.843pt}{0.117pt}}
\multiput(1337.00,788.17)(11.251,7.000){2}{\rule{0.421pt}{0.400pt}}
\multiput(1350.00,796.59)(0.824,0.488){13}{\rule{0.750pt}{0.117pt}}
\multiput(1350.00,795.17)(11.443,8.000){2}{\rule{0.375pt}{0.400pt}}
\multiput(1363.00,804.59)(0.874,0.485){11}{\rule{0.786pt}{0.117pt}}
\multiput(1363.00,803.17)(10.369,7.000){2}{\rule{0.393pt}{0.400pt}}
\multiput(1375.00,811.59)(0.950,0.485){11}{\rule{0.843pt}{0.117pt}}
\multiput(1375.00,810.17)(11.251,7.000){2}{\rule{0.421pt}{0.400pt}}
\multiput(1388.00,818.59)(0.824,0.488){13}{\rule{0.750pt}{0.117pt}}
\multiput(1388.00,817.17)(11.443,8.000){2}{\rule{0.375pt}{0.400pt}}
\multiput(1401.00,826.59)(0.950,0.485){11}{\rule{0.843pt}{0.117pt}}
\multiput(1401.00,825.17)(11.251,7.000){2}{\rule{0.421pt}{0.400pt}}
\multiput(1414.00,833.59)(0.874,0.485){11}{\rule{0.786pt}{0.117pt}}
\multiput(1414.00,832.17)(10.369,7.000){2}{\rule{0.393pt}{0.400pt}}
\multiput(1426.00,840.59)(0.824,0.488){13}{\rule{0.750pt}{0.117pt}}
\multiput(1426.00,839.17)(11.443,8.000){2}{\rule{0.375pt}{0.400pt}}
\end{picture}

\vskip 0.2cm
{\small \noindent Figure 5: The measured effective size, $R_p$, of
the transiting planets (from top to bottom) 
HD209458b, OGLE-TR-56b and OGLE-TR-113b
versus $\sqrt{T_s/M_P}$ (in units where $\sqrt{T_s/M_P} = 1$
for HD209458b). The solid line
is the prediction, Eq.(\ref{1new}), which assumes that
the planets are 
composed predominately of mirror matter.}
\vskip 1.5cm
Evidently the 2001 prediction, Eq.(\ref{1new}), is in reasonable 
agreement with the observations.
This appears to be non-trivial: in the case of ordinary matter
planets, increasing the mass does not significantly affect
the radius, and does not generally lead to a decreasing
radius (for example, Jupiter is three times heavier than Saturn,
but is 15\% {\it larger}).
However, it is possible that the apparent agreement
with the rough prediction, Eq.(\ref{1new}) is coincidental -- so
more data would be welcome.
Especially decisive would be
the discovery of a much heavier transiting planet,
$M_P \stackrel{>}{\sim} 2M_J$,
which should have a radius less than $R_J$ if it is
a mirror world.

\vskip 0.2cm

\subsection{Isolated planets?}
\vskip 0.2cm

If this mirror world interpretation of the close-in planets
is correct then it is very natural that the dynamical
mirror image system of a mirror star with an ordinary planet
could also exist. Such a system would appear to ordinary observers
as an ``isolated" ordinary planet. Remarkably,
such ``isolated" planets have been identified
in the $\sigma$ Orionis star cluster\cite{iso}.
These planets have estimated mass of $5-15 M_{Jupiter}$
and appear to be gas giants which do not seem
to be associated with any visible star. 
Given that the $\sigma$ Orionis cluster is estimated to be less than  
5 million years old, the formation of these ``isolated" planets
must have occurred within this time (which
means they can't orbit faint stellar bodies such
as old white dwarfs). Zapatero Osorio et al\cite{iso}
argue that these findings pose a challenge to
conventional theories of planet formation which are
unable to explain the existence of numerous isolated planetary mass
objects. Thus, the existence of these planets is surprising 
if they are made of ordinary matter,
however there existence is natural from the mirror world
perspective since they can be interpreted as ordinary planets orbiting
mirror stars\cite{iso2}. 
Furthermore, if the isolated planets
are not isolated but orbit mirror stars then
there must exist a periodic Doppler shift detectable
on the spectral lines from these planets. This
represents a simple way of testing this hypothesis\cite{iso2}.

\vskip 0.2cm
\subsection{Anomalous impact events}
\vskip 0.2cm

Perhaps the most fascinating possible implication of
mirror matter-type dark matter is that
our solar system contains mirror matter
space-bodies (SB)\cite{sil,tung}. 

There is not much room for a large amount of mirror 
matter in our solar system. For example, the amount of mirror matter 
within the Earth has been constrained to be less than $10^{-3}
M_{Earth}$\cite{sashaV}. However, we don't know enough about the
formation of the solar system to be able to exclude the existence
of a large number of space bodies made of mirror matter if
they are small like comets and asteroids. The total mass of
asteroids in the asteroid belt is estimated to be only about
0.05\% of the mass of the Earth. A similar or even greater number
of mirror bodies, perhaps orbiting in a different plane or even
spherically distributed like the Oort cloud is a fascinating 
possibility\footnote{ Large planetary sized
bodies are also possible if they are in distant
orbits\cite{sil} or masquerade as ordinary planets or moons
by accreting ordinary matter onto
their surfaces.}. In fact, the comets themselves -- and hence
the Oort cloud itself -- might actually
be composed of mirror matter (as we will discuss in the following
subsection).

Anyway, collisions of such bodies with themselves and
ordinary bodies would generate a solar system population of
mirror gas and dust particles and larger bodies.
The impact velocity of such
solar system objects (relative to the Earth) would
be in the range\footnote{
The minimum velocity is the result of the local
acceleration of the Earth (equivalent to the
escape velocity for a particle on the Earth, which is 11.2 km/s).}:
\begin{eqnarray}
11\ {\rm km/s} \stackrel{<}{\sim} v \stackrel{<}{\sim} 70 \ {\rm km/s} \ .
\end{eqnarray}
If such small mirror matter bodies do in fact exist and happen to collide
with the Earth, what would be the consequences? If the only force connecting
mirror matter with ordinary matter is gravity, then the consequences would
be minimal. The mirror matter space body would simply pass
through the Earth and nobody would know about it unless
the body was so heavy as to gravitationally affect the
motion of the Earth. However, if there is photon-mirror
photon kinetic mixing of magnitude, $\epsilon \sim
5\times 10^{-9}$, as indicated by the DAMA/NaI experiment, then the
mirror nuclei of the space body can interact with the ordinary nuclei 
in the Earth via elastic Rutherford scattering (see Figure 2).

Small dust particles could thereby be detectable in
simple surface experiments. In particular, experiments such
as the St. Petersburg experiment\cite{drob} are sensitive to
solar system mirror dust particles\cite{sm03}. Such particles can
produce a burst of photons in a scintillator due to
elastic collisions between the mirror atoms of the dust particle
and the ordinary scintillator atoms.
Not only can these photons be detected via
a photomultiplier (PM) tube,
but the velocity of the mirror dust particle can be determined 
if the PM tubes are appropriately arranged.
This is important because
ordinary cosmic rays should be travelling close to the
speed of light, and can thereby be distinguished from
relatively slow moving mirror dust particles.
The St. Petersburg experiment finds a positive signal consistent
with a flux of about 1 mirror dust particle per square meter per day.

Impacts of larger bodies should be less frequent, nevertheless there
is a fascinating range of evidence for their existence.
The largest recorded impact event was the 1908 Tunguska
event. Remarkably no significant asteroid or cometary remnants 
were recovered from the Tunguska site\cite{rev}.\footnote{
There is some interesting evidence for
microscopic particles in tree resin\cite{resin},
which might have originated from the Tunguska space-body.
However their tiny abundance is hardly consistent
with reasonable expectations.} People have
{\it assumed} that the impacting body was made of ordinary
matter, however there is (literally!) no solid evidence to support
this claim.
The Tunguska body may have been made
out of dark matter -- which is a logical possibility
if mirror matter is identified with the dark 
matter of the Universe. In fact, this hypothesis seems
to provide a better explanation for the known features
of the Tunguska event\cite{tung}.
There are also many other `anomalous' impact events, on smaller
scales\cite{small}, and evidence for anomalous impact events on
larger scales. Included among the latter are the impact events 
responsible for strange glass 
fields such as Edeowie glass\cite{haines},  
Libyan desert glass, tektites etc.
All of these impact related phenomena share a 
common feature which is the remarkable lack
of clearly defined extraterrestrial material 
or even chemical traces (such as 
iridium excess). This fact is obviously explicable if the
events were due to the impact of a mirror matter
space body.

Other solar system evidence for mirror matter
also exists coming from the lack of
small craters on the asteroid EROS\cite{eros,s1} and also
from the anomalous slow-down of {\it both}
Pioneer spacecraft\cite{study,fvpioneer}. 
The overall situation is
summarized in figure 6. 
\vskip 0.4cm
\centerline{\epsfig{file=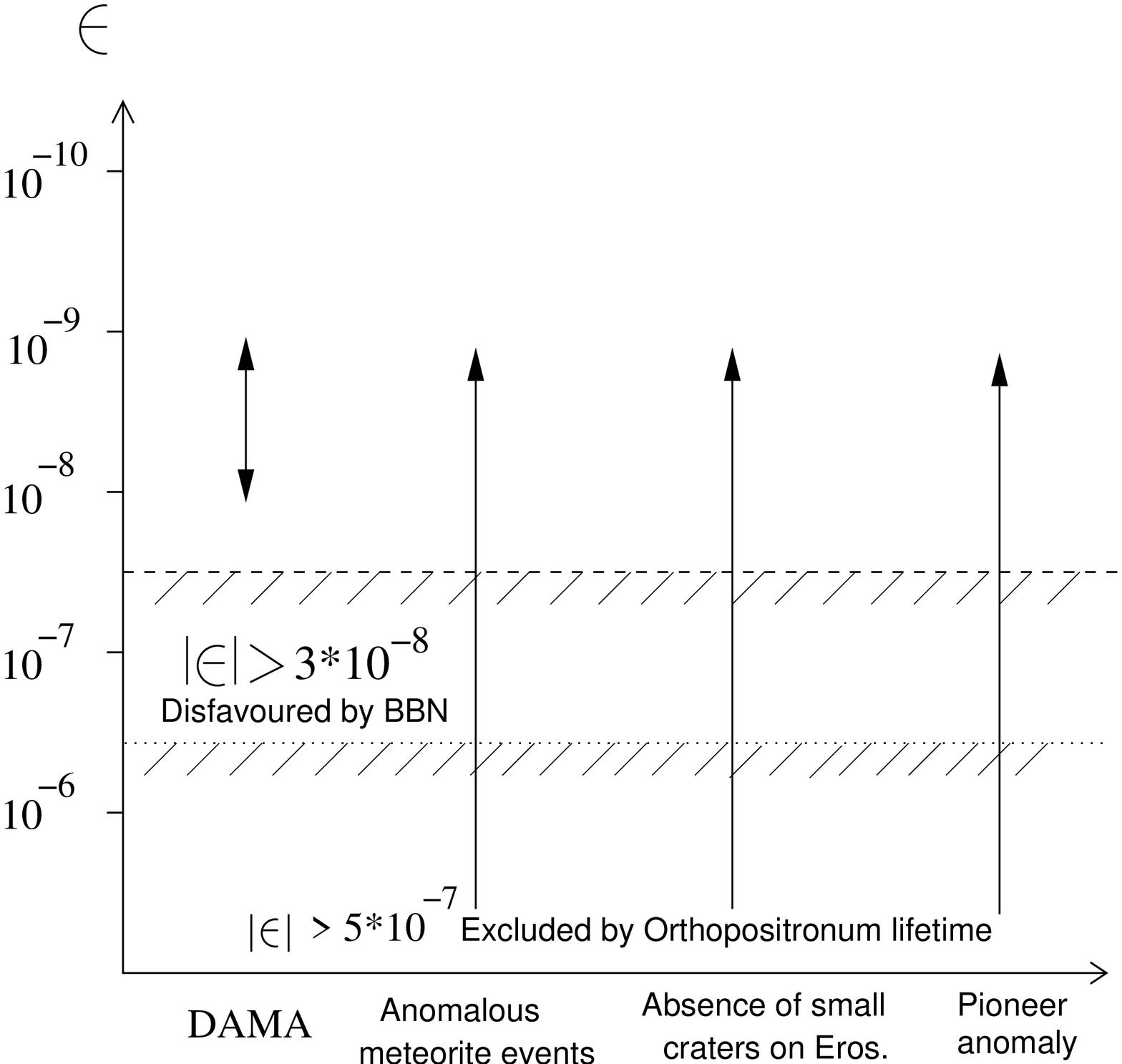,height=6.8cm,width=9.5cm}}
\vskip 0.3cm
\noindent
{\small Figure 6: Favoured range of $\epsilon$ from 
various experiments/observations. Also shown are the
current direct experimental bound, $\epsilon < 5 \times 10^{-7},$
which comes from orthopositronium lifetime
studies\cite{fg,ortho} and also the limit, $\epsilon < 3\times 10^{-8}$,
suggested by early Universe cosmology (successful BBN) \cite{cg}.}
\vskip 1.3cm

Direct detection of mirror matter fragments in the ground is also
possible at
these impact sites. The 
photon-mirror photon kinetic mixing interaction will lead to a 
static force which can keep small mirror matter fragments
(of size $R$) near the Earth's surface, provided
that\cite{cent}
\begin{eqnarray}
R \stackrel{<}{\sim} few \left({|\epsilon | \over 5\times
10^{-9}}\right)\ {\rm cm.}
\end{eqnarray}
Such fragments can be experimentally detected via
the centrifuge technique\cite{cent}
and through the thermal effects of the embedded mirror matter on
the surrounding ordinary matter\cite{th6}.
Note however that impacts of galactic halo 
mirror ions/electrons will vaporize these small fragments over time.
The flux of halo mirror electrons is roughly
$f_h \sim 10^8 \ {\rm cm^{-2} s^{-1}}$
and defining $X$ to be the mean number of mirror atoms evaporated
from the impact of each halo mirror electron ($X \sim 10$), then the 
rate at which a mirror matter fragment would evaporate
would be of order $dR/dt = f_h X/n \sim 1 \ {\rm cm/Myr}$
(where $n \sim 10^{23}{\rm /cm^3}$ is the atomic number density 
of mirror atoms in the mirror fragment). This suggests that
mirror matter fragments probably could not be recovered from
the remnants of
old impact events, such as Edeowie glass, Libyan desert glass, tektites etc
(which are of order 1 Myr old or older)
but might be recovered from relatively recent
anomalous impact events such as the Tunguska event\cite{rev} and 
small anomalous impact events\cite{small}.

If mirror matter space-bodes do exist in our 
solar system, then one might expect other
unconventional scientific implications. Below we mention just a few
more of these things.

\subsection{Are comets made of mirror matter?}

Comets are believed to originate from an
approximately spherically symmetric cloud extending out
about half way to the nearest star. This comet cloud, called
the Oort cloud, is reminisant of the dark halo of our galaxy.
Both are largely invisible, are distributed differently to
the `visible' matter, and are also hypothetical. Of course,
this analogy is very simplistic and should not be taken very
seriously. Nevertheless, it is also true that
comets seem to have a number of puzzling features and
are not altogether well understood.
One interesting feature of comets is that 
they seem to contain a very dark nucleus.
%as shown in the picture of comet Halley.
%\newpage .
%\vskip -3.2cm
%\vskip 3cm
%\centerline{{\Large Figure halley.gif}}
%\vskip 1cm
%\centerline{\epsfig{file=fig10.eps,width=12.2cm}}
%\noindent
%{\small  Comet Halley's Nucleus. This picture
%was taken by  the spacecraft Giotto. Contrary to prior
%expectations, Halley's nucleus is very dark -- one
%of the darkest objects in the solar system. Credit:
%Halley Multicolour Camera Team, ESA.}
%\vskip 1.2cm
For example, the
nucleus of Halley's comet has an albedo of only 0.03 making it one
of the darkest objects in the solar system -- darker
even than coal! This has led to the suggestion\cite{sug,sug2} that
the nucleus could be composed predominately of mirror
matter. Of course, pure mirror matter would be
transparent, but if it contained a small admixture
of ordinary matter embedded within, it might appear
opaque and dark. If the ordinary matter had a volatile
component such as water ice, then this would explain
the large head and tail observed when the comet passed
close to the sun. 
Furthermore, such a picture would 
simply explain the long standing comet fading problem:
that many comets lose a large factor 
(100-1000) in average brightness
after approaching the sun for the first time.
If this interpretation is correct, then comets may
simply become dimmer and dimmer over time rapidly
losing all of their volatile ordinary matter component.
They may effectively become invisible. Of course, the
rate that this occurs will depend on many things
such as the proportion of ordinary to mirror matter,
the chemical composition, details of the orbit etc.

Interestingly, a recent study\cite{yyy} has concluded
that many old comets must have either become invisible
or have somehow disintegrated.
The number of cometary remnants (assumed to be asteroid-like objects) is
100 times less abundant than theoretically expected!\cite{yyy}.
Clearly, this seems to support (or at least, encourage) the 
mirror matter interpretation of the comets. 
Of course, if comets are predominately made 
of mirror matter then this fits-in nicely with the
mirror matter interpretation of the anomalous small impact events
(and Tunguska event), which was discussed in the previous
subsection.
It might also be connected with atmospheric anomalies. 

\subsection{Atmospheric anomalies caused by small mirror 
matter space-bodies?}

To explain the anomalous small impact events 
we require that some of the mirror matter space-bodies to
survive and hit the ground without completely melting and
vaporizing in the atmosphere.  Detailed studies\cite{tung}
have shown that this is possible, 
especially for non-volatile
mirror matter (such as mirror iron).
Sometimes, it could happen that a mirror space-body
would heat up enough to completely vaporize
in the atmosphere.
After vaporizing, the mirror atoms interact
with the air atoms by Rutherford scattering. Although
initially the mirror matter will heat up the ordinary matter
because of its large kinetic energy (since its initial
velocity is at least 11 km/s), after a short time, the mirror matter
will cool the atmosphere. The mirror atoms will draw in heat
from the surrounding ordinary atoms and radiate it away into
mirror photons. Since the mirror atoms are not absorbing mirror
photons from the environment, heat will be lost from the system.
The net effect is a localized rapid cooling 
of the atmosphere which might lead to the formation of unusual clouds
and other strange atmospheric phenomena.
This might
explain the remarkable observations of
falling ice blocks\cite{ib} and maybe even the observations
of atmospheric `holes'\cite{hole}.
It seems that the answer may indeed be `blowing in the wind' --
but only for a sort time!
\footnote{
Eventually the mirror atoms will disperse and ultimately be
evaporated from the Earth due to interactions with halo 
mirror atoms.} 

\section{Conclusion}

Historically, imposing symmetries of particle interactions
has led to the prediction and subsequent discovery of 
a variety of `new' fundamental particles including:
\begin{itemize}
\item Antiparticles -- predicted to exist by imposing proper
Lorentz symmetry;
\item Neutrino -- predicted to exist by imposing time translational 
symmetry (energy conservation);
\item Top quark -- predicted to exist from $SU(2)\otimes U(1)_Y$ 
electroweak gauge symmetry (to partner the bottom quark);
\item The $\Omega^-$ baryon -- predicted from $SU(3)$ flavour
symmetry in the quark model.
\end{itemize}
Mirror matter is also an offspring of this
methodology; it is an attempt to follow this historically 
successful approach. In fact it appears to be 
theoretically unique,
arising from a single well motivated hypothesis: The improper
Lorentz symmetries (such as parity and time reversal invariance)
stand out as the
only space-time symmetries which are not
respected by the interactions of the known
elementary particles, but can be exact unbroken symmetries
of nature if a set of mirror particles exist.

Mirror matter is thus 
very well motivated from a particle physics point of view. 
Furthermore it seems to have the right properties to be identified with 
the inferred non-baryonic dark matter in the Universe. Specifically,
mirror dark matter
seems to provide a consistent explanation for: a) the 
basic dark matter
particle properties (mass, stability, darkness); b)
the similarity in cosmic abundance between ordinary
and non-baryonic dark matter, $\Omega_B \sim \Omega_{dark}$;
c) large scale structure formation;
d) microlensing (MACHO) events; e) asymptotically
flat rotation curves in spiral galaxies
and f) the impressive DAMA/NaI annual modulation signal.

Of course, any theory of dark matter should also be measured
against the standard paradigm -- that non-baryonic dark matter 
consists of hypothetical weakly interacting particles i.e.
essentially collisionless particles. However
this comparison is actually favourable. In the standard 
WIMP hypothesis:
the basic dark matter properties (stability, darkness) require
ad hoc hypothesis; MACHO events cannot be explained; $\Omega_{dark}\sim
\Omega_B$ is mysterious; DAMA/NaI annual modulation signal is difficult
to understand consistently with 
other experiments such as CDMSII.
Perhaps the only thing that WIMPs might explain better is the 
existence of the halo in galaxies.
This is because WIMPs are non-dissipative. However this
success is significantly eroded by the facts;
standard collisionless dark matter predicts\cite{fin1} overly dense
cores in galaxies and
over abundance of small scale structures within halos
which are {\it not} consistent
with the observations\cite{salucci}. 

Thus, by either comparing 
mirror matter-type dark matter
with experiments and observations 
or with the standard WIMP paradigm, it is clear that it is
a strong candidate for the non-baryonic dark matter in
the Universe, deserving of serious consideration and further study.

\vskip 0.4cm
\noindent
{\large \bf Acknowledgements}
\vskip 0.1cm
\noindent
It is a pleasure to thank my collaborators,
Sergei Gninenko, Sasha Ignatiev,
Henry Lew, Saibal Mitra, Zurab Silagadze, Ray Volkas and T. L. Yoon.
I would also like to thank S. Mitra, Z. Silagadze and R. Volkas for
their comments on a draft of this article.
This work was supported by the Australian Research Council.

\end{document}